\documentclass[10pt,conference,a4paper]{IEEEtran}
\IEEEoverridecommandlockouts

\usepackage{cite}
\usepackage{amsmath,amssymb,amsfonts}
\usepackage{textcomp}

\usepackage[caption=false]{subfig}
\usepackage{booktabs}
\usepackage{url}
\usepackage{hyperref}
\usepackage{bm}
\usepackage{mathtools}

\usepackage{array}

\DeclarePairedDelimiter\autobracket{(}{)}
\newcommand{\br}[1]{\autobracket*{#1}}
\newcommand{\bmx}{\bm{x}}
\newcommand{\bmy}{\bm{y}}
\newcommand{\bmt}{\bm{\theta}}
\newcommand{\bmm}{\bm{\mu}}
\newcommand{\bms}{\bm{\sigma}}
\newcommand{\bmss}{\mathbf{s}}

\newcommand{\grad}{\hat{\nabla}}
\newcommand{\prior}{p\br{\bm{\theta}}}
\newcommand{\vpost}{q\br{\bm{\theta}}}

\newcommand{\Dcal}{\mathcal{D}}
\newcommand{\Lcal}{\mathcal{L}}

\newcommand{\eq}{\,{=}\,}

\def\BibTeX{{\rm B\kern-.05em{\sc i\kern-.025em b}\kern-.08em
    T\kern-.1667em\lower.7ex\hbox{E}\kern-.125emX}}
    
\usepackage{soul}
\usepackage{float}

\begin{document}

\title{Bayesian Bilinear Neural Network for Predicting the Mid-price Dynamics in Limit-Order Book Markets}

\author{\IEEEauthorblockN{Martin Magris}
\IEEEauthorblockA{Department of Electrical and\\ Computing Engineering \\
Aarhus University\\
Denmark\\
Email: magris@ece.au.dk}
\and
\IEEEauthorblockN{Mostafa Shabani}
\IEEEauthorblockA{Department of Electrical and\\ Computing Engineering \\
Aarhus University\\
Denmark\\
Email: mshabani@ece.au.dk}
\and
\IEEEauthorblockN{Alexandros Iosifidis}
\IEEEauthorblockA{Department of Electrical and\\ Computing Engineering \\
Aarhus University\\
Denmark\\
Email: ai@ece.au.dk}}

\maketitle

\begin{abstract}
The prediction of financial markets is a challenging yet important task. In modern electronically-driven markets, traditional time-series econometric methods often appear incapable of capturing the true complexity of the multi-level interactions driving the price dynamics. While recent research has established the effectiveness of traditional machine learning (ML) models in financial applications, their intrinsic inability to deal with uncertainties, which is a great concern in econometrics research and real business applications, constitutes a major drawback. Bayesian methods naturally appear as a suitable remedy conveying the predictive ability of ML methods with the probabilistically-oriented practice of econometric research. 
By adopting a state-of-the-art second-order optimization algorithm, we train a Bayesian bilinear neural network with temporal attention, suitable for the challenging time-series task of predicting mid-price movements in ultra-high-frequency limit-order book markets. 
We thoroughly compare our Bayesian model with traditional ML alternatives by addressing the use of predictive distributions to analyze errors and uncertainties associated with the estimated parameters and model forecasts. Our results underline the feasibility of the Bayesian deep-learning approach and its predictive and decisional advantages in complex econometric tasks, prompting future research in this direction.
\end{abstract}


\section{Introduction}
Bayesian inference is known to be a difficult task outside a relatively small class of well-studied models, generally involving conjugate priors for the likelihood. The analytical Bayesian treatment of general, even small-dimensional, problems is widely unfeasible. The increased computational capacity available these days, as much as the availability of powerful algorithms such as Monte Carlo Markov Chain or Metropolis-Hastings, opened the possibility for a simulation-based approach to Bayesian inference. However, Bayesian methods in typical large-scale complex Machine Learning (ML) problems have long been impractical. Though ML generally operates under a frequentist perspective, the first steps into a probabilistic approach to Deep Learning (DL) are relatively recent, see e.g., \cite{murphy_machine_2012,gal_dropout_2016} and references therein. Only recently, we are witnessing a growing interest in Bayesian DL, boosted by its demand across multiple disciplines. Indeed, even in a simulation-aided setting, Bayesian inference on the potentially thousands of parameters over highly non-linear models like Neural Networks (NN) is certainly not a simple task.

Yet, the interest in probabilistic modeling and Bayesian methods in other disciplines has a much longer history. Especially in econometrics and finance, the probabilistic dimension is an innate and essential element in modeling. Indeed econometric research is at the cross-edge between applied statistics, probability theory, stochastics, and the study of economic phenomena \cite{frisch1933}. As such, the econometric practice is that of developing well-reasoned and economically motivated, essential, mostly parametric, probabilistic models that are thoroughly tested, validated and back-tested following the principles of statistical inference. For example, the concepts of significance testing, confidence intervals, asymptotic analysis, and stationarity are typical in the econometric literature. On the other hand, such an approach in DL is currently inapplicable. 

At the same time, researchers in economics and practitioners in finance acknowledge the flexibility, scalability, and gains in predictive tasks that ML can bring when applied to economic problems, e.g., \cite{mullainathan2017machine,dixon2020machine}. Especially in modern, electronically-driven financial markets, operating at ultra-high frequencies and generating massive complex and multidimensional datasets underlying the complex dynamics of market variables arising from the interactions of multiple players and forces at different levels, ML methods have gained much attention, see, e.g., \cite{varian2014big}. Business and financial applications are part of those high-risk domains where quantifying the uncertainty underlying models' estimates and predictions is of utmost importance \cite{salinas2020deepar}. A probabilistic dimension reflecting uncertainties related to model estimation and perhaps accounting for the typical elements of business activity that are difficult to predict \cite{makridakis2009forecasting}, would be beneficial.

The recent advances in Bayesian DL have the potential of bringing this element into play, narrowing the gap between the highly-probabilistic yet parsimonious modeling of the econometric practice and the flexible non-linear and non-parametric ML rationale. 
Bayesian inference for neural networks has recently been shown to be challenging yet feasible \cite{kingma_auto_2014,osawa_practical_2019,blundell_weight_2015}. Bayesian Neural Networks (BNNs) are engaged with the typical elements of Bayesian inference, in particular with a trainable distribution over its parameters and a consequent predictive distribution that enables classical statistical tools, econometric methods, and relevant risk-related and uncertainly-related analyses, e.g., based on predictive distributions \cite{geweke2010comparing}. Nevertheless, much research in this direction is still needed. 

This paper aims to introduce the use of BNNs in economic problems in light of the above discussion, boosting further research and interest in this research direction. 
We propose a Bayesian version of the Temporal Attention-augmented Bilinear network as a lightweight DL model for a financial times-series classification task. We propose a first Bayesian DL econometric application in the challenging task of predicting mid-price movements in Limit Order Book (LOB) markets. Our results explore the feasibility of such an approach, compare its forecasting performance against non-Bayesian specifications based on different optimization algorithms, and address the advantages of adopting BNNs in financial applications.

\section{Literature review}
Neural Networks (NNs) have been successfully applied in several ML problems, such as image classification \cite{krizhevsky_imagenet_2012,simonyan_very_2015} computer vision \cite{girshick_rich_2014,ren_faster_2015}, natural language processing \cite{collobert_unified_2008,goldberg_neural_2017} or speech recognition \cite{mohamed_acoustic_2012,dahl_context-dependent_2012}. Despite their undeniable use and performance in delivering leading results on different predictive tasks, the decisions are achieved in a rather uninterpretable manner. 

NNs correspond to statistical black-box models that achieve feasible point estimate predictions by adapting their high-dimensional parameters on a try-and-error basis. Based on the nature of the problem, the user defines a cost function and a network architecture that allows to approximate complex non-linear functions to tackle a prediction or classification task.
By allowing for a sufficiently large number of trainable parameters, under mild assumptions, NNs can approximate any arbitrary function \cite{cybenko_approximation_1989,Lu2017expressive,Hanin2019universal}. It is with little surprise that NNs found significant use in financial and econometrics applications, where complex and interacting latent structures in the data drive the behavior of different economic variables. A review of different NN applications in finance is provided in \cite{mcnelis_neural_2005}, an early discussion on econometric applications can be found in \cite{kuan_artificial_1994} and within time-series analysis in \cite{terasvirta_linear_2005,m_qi_trend_2008,hewamalage2021recurrent}.
\cite{cenesizoglu_effects_2014} analyzed the relationship between LOB variables and mid-price movements showing that it is possible to obtain economical gain from these variables and the mid-price return. Further, their causality analysis supports the use of lagged LOB variables for forecasting purposes. The design of a set of features, extending the LOB feature set in the high-frequency forecasting application of \cite{kercheval_modelling_2015}, can be found in \cite{ntakaris_feature_2019}, while \cite{ntakaris_benchmark_2018,ntakaris_feature_2019} tackle the mid-price movement prediction with the rich LOB data under different ML perspectives, including NNs. For a similar prediction task, \cite{dixon_sequence_2018} addressed the use of recurrent NN. The use of Long-Short-Term Memory (LSTM) networks and Convolutional Neural Networks (CNNs) is discussed in \cite{tsantekidis_forecasting_2017,zhang2019deeplob,passalis2019adaptive,tsantekidis_using_2020} and the use of Neural Bag of Features in \cite{passalis2017time,passalis2018temporal}. Taking advantage of the spatial structure in the LOB, \cite{sirignano_deep_2019} provides an extensive analysis of over 500 stocks for prediction price movements, while \cite{tran2017tensor} encodes the LOB data as two-order tensors. An attention mechanism capable of exploiting and retaining the temporal mode of the order flow is introduced in \cite{tran_temporal_2019} and extended to accommodate multiple attentions in \cite{shabani_multi-head_2022}.

By fine-tuning the network and increasing the number of parameters, one possibly achieves functions with higher complexity and improved forecasting ability. The lack of interpretation and the impossibility of condensing the decision process to a simple decision rule, along with overfitting issues and their native non-probabilistic setup, create challenges in the use of NNs in high-risk domains and for all those applications where uncertainties in predictions are of relevance \cite{goan_bayesian_2020}. Moreover, the lack of a well-defined building protocol (e.g., the absence of an Akaike Information Criterion -like statistic for features' relevance determination) makes their adoption by experts and practitioners from such domains difficult \cite{caruana_intelligible_2015,holzinger_what_2017,vu_shared_2018,holzinger_causability_2019}.

A Bayesian perspective on NNs provides a natural way to reason around uncertainties. At the same time, it provides tools for model regularization and offers insights into how decisions are made. Indeed, the Bayesian paradigm offers a perspective on NNs that can address many of the issues currently faced by NNs. 
Recent research investigated how Bayesian principles can adapt to large NNs. To this end, a learnable distribution is placed over the parameters, resulting in BNNs.
A survey on early developments in BNNs can be found in \cite{mackay_probable_1995}, and recent introductions to BNNs are those of, e.g.,\cite{lampinen_bayesian_2001,jospin_hands_2020,goan_bayesian_2020}. For a specialized survey on algorithms for training BNNs see \cite{magris2023}.
In BNNs, parameters are treated as random variables, and the learning focuses on the distribution of these parameters conditional on the observed training data sample. In the learning phase, the latent distribution of the parameters is inferred based on the current knowledge and the observed data by use of the Bayes theorem resulting in a distribution of the model parameters conditional on the data, the posterior distribution. Further details on BNNs and their training appear in Section \ref{sec:methods}.

Financial applications involving BNNs are somewhat limited.
An application for automatic relevance determination in option pricing is that of \cite{mbuvha_automatic_2019}. A recent example of stock-price prediction is found in \cite{chandra_bayesian_2021}, where exploitative MCMC-based learning is used to forecast daily closing prices of four stocks, showing that in terms of RMSE performance metric, their BNN outperforms non-Bayesian counterparts. A forecasting study based on electricity prices is provided in \cite{vahidinasab_bayesian_2008,ghayekhloo_combination_2019}, while Bitcoin data is used in \cite{jang_empirical_2018}. Sign-changes in returns have been analyzed under a Multilayer Perceptron (MLP) BNN in \cite{skabar_direction-change_2009}. This study uses low-frequency daily closing prices and lagged moving averages as features, showing a slight 52\% accuracy over a random classifier and no gains with respect to a standard MLP. There are, however, no applications involving tick-by-tick data generated from typical modern financial markets running over the LOB systems. 
Nevertheless, the bridging potential that BNNs could provide between the fields of econometrics and ML has not been recognized.

\section{Methods} \label{sec:methods}
\subsection{Bayesian Neural Networks} 

A BNN is any stochastic Artificial Neural Network (ANN) trained using Bayesian inference. ANNs aim at approximating arbitrary functions $\bmy \eq NN_{\theta}\br{\bmx}$, whose parameters are denoted by $\theta$. Over a training dataset $\Dcal$, the standard estimation approach is to determine a minimal-cost point estimate $\hat{\theta}$ using backpropagation. In BNNs, parameters are treated as latent random variables, and the goal is to learn the distribution of the parameters conditional on $\Dcal$. The first step is that of defining the joint distribution of the data and the parameters $p\br{\theta,\Dcal}=p\br{\Dcal|\theta}p\br{\theta}$, which depends on our prior belief over the latent variables $p\br{\theta}$ and the chosen form of likelihood $p\br{\Dcal|\theta}$. Under independence between the model parameters and the inputs, the Bayesian posterior is written as $p\br{\theta|\Dcal} = p\br{\Dcal, \theta} p\br{\theta} / p\br{\Dcal}$. Computing the weight-independent term known as marginal likelihood (or evidence) is perhaps one of the most difficult tasks in Bayesian inference: the prior-to-posterior update is usually intractable.
From the posterior distribution, the model uncertainty is quantified as the marginal probability distribution of the output $\bmy_i$ for a certain input $\bmx_i$, through the predictive distribution
\begin{equation}\label{eq:predictive}
    p\br{\bmy_i|\bmx_i,\Dcal} = \int p\br{\bmy_i|\bmx_i,\theta} p\br{\theta | \Dcal} d\theta.
\end{equation}
When performing classification, the average model prediction approximates the relative probability of each class, 
\begin{equation}\label{eq:bayesian_model_averaging}
\hat{p}_{ic} \approx 1/N_s \textstyle\sum_{n=1}^{N_s} p\br{y_i = c| \bmx_i, \theta^{\br{n}}} \text{,}
\end{equation}
where $\theta^{\br{n}} {\sim}\, p\br{\theta | \Dcal}$.
If the cost of giving a false positive is equal across all classes, the final classification is taken according to the most likely class, i.e.,
\begin{equation}\label{eq:classification_crit}
\hat{y}_i = \max_c \hat{p}_{ic}.
\end{equation}

\subsection{Temporal Attention-augmented Bilinear network} \label{sec:TABL} 
The Temporal Attention-augmented Bilinear Network (TABL) architecture \cite{tran_temporal_2019} is a lightweight DL model which has been shown to be particularly suited for multidimensional time-series forecasting. It augments the bilinear projection with an attention mechanism exploiting the temporal dimension across the features. This enables it to compare favorably with alternative architectures such as Bilinear networks, CNNs, LSTM networks, and several other ML algorithms \cite{tran_temporal_2019}.

\begin{figure}[ht]
\centering
\scalebox{0.8}{
\includegraphics[width = 0.55\textwidth]{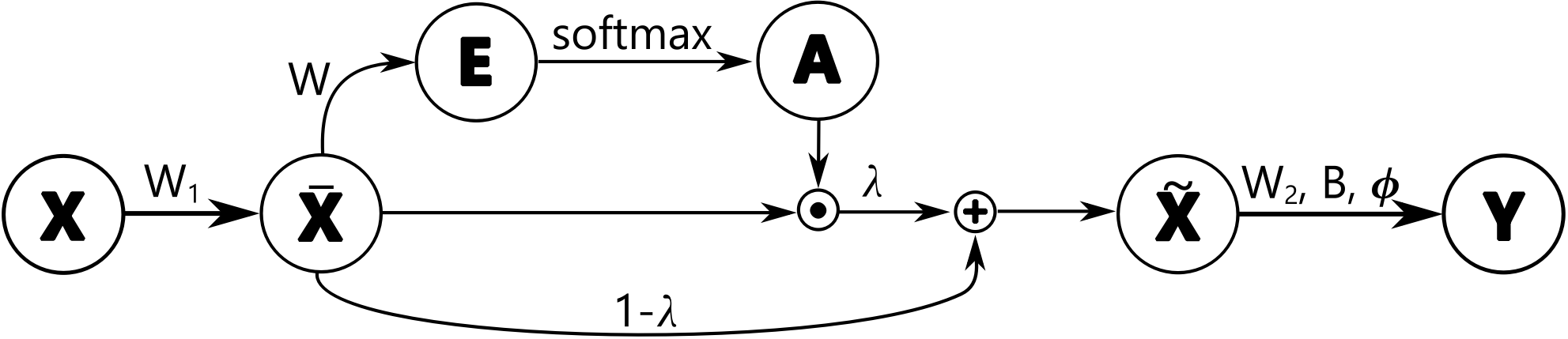}
}
\caption{Illustration of the TABL architecture.}
\label{fig:TABL}
\end{figure}  

Figure \ref{fig:TABL} illustrates the architecture of the TABL layer. It maps a $D\times T$-dimensional input matrix  $\mathbf{X}$ onto a $D' \times T'$-dimensional output $\mathbf{Y}$, where $D$ and $D'$ correspond to the number of features and $T$, $T'$ correspond to the number of temporal instances. 
The network initially operates a projection of the temporal dimension of the input matrix to a $D'\times T$ -dimensional feature space modeling the dependence on the first mode while preserving the temporal order of the features. It further learns the relative importance of the temporal instances with respect to each other, producing an attention mask where only the most relevant instances are preserved. A learnable scalar drives the mixture of the temporal and non-temporal features passed to a final mapping that returns the final representation adjusted for bias. 
This is achieved by:
\begin{align}
    \bar{\mathbf{X}} &= \mathbf{W}_1 \mathbf{X}\\
    \mathbf{E} &= \bar{\mathbf{X}} \mathbf{W}\\
    a_{ij} &=  \exp(e_{ij}) / {\textstyle\sum_{k=1}^T} \exp(e_{ik}) \\
    \tilde{\mathbf{X}} &= \lambda (\bar{\mathbf{X}} \odot \mathbf{A}) + (1-\lambda) \bar{\mathbf{X}}\\
    \mathbf{Y} &= \phi ( \tilde{\mathbf{X}} \mathbf{W}_2 + \mathbf{B})
\end{align}
where $a_{ij}$ and $e_{ij}$ denote the element $(i,j)$ of $\mathbf{A}$ and $\mathbf{E}$, respectively, $\phi(\cdot)$ is a predefined activation function, $\mathbf{W}_1 \in \mathbb{R}^{D'\times D}$, $\mathbf{W} \in \mathbb{R}^{T\times T}$, $\mathbf{W}_2 \in \mathbb{R}^{T\times T'}$ and $\mathbf{B} \in \mathbb{R}^{D' \times T'}$ are the parameters of the layer, and ${0  \leq \lambda \leq 1}$ is a learnable mixing coefficient which determines the importance of using the temporal attention in the mapping. 
Experiments in \cite{shabani_multi-head_2022} further show that the inclusion of additional temporal attention heads opens toward richer structures in the temporal dependence across lagged features, relevant for forecasting purposes.

\subsection{Bayesian Temporal Attention-augmented Bilinear network}\label{sec:VOGN} 
To formulate the Bayesian network formed by one TABL layer (B-TABL), we define the parameter vector $\bm{\theta}$ formed by the parameters of TABL, i.e., $\bmt = \{\mathbf{W}, \mathbf{W}_1, \mathbf{W}_2, \mathbf{B}, \lambda\}$.
Variational Inference (VI)  is a well-established methodology for complex statistical inference circumventing the typical intractable integration problem arising in Bayesian inference by approximating the true posterior $p\br{\bm{\theta} | \Dcal}$ with a distribution $\vpost$ whose normalization constant is easier to compute (variational distribution). A review on VI from a statistician's perspective is that of \cite{blei2017variational}, from the ML perspective, that of \cite{tran_practical_2021}, while very recent applications in multidimensional econometric models are, e.g., those of \cite{gunawan2021variational,gefang2022}. Fixed form variational Bayes assumes a fixed parametric form for the density in some class of distributions $\mathcal{Q}$, indexed by a variational parameter vector. A perspective on the problem with general non-conjugate likelihoods for priors in the exponential family can be found in \cite{khan_2018_fast}. We chose both $\prior$ and $\vpost$ to be Gaussian distributions with diagonal covariance matrices:
\begin{equation*} 
   \prior = \mathcal{N}\br{\theta|\bm{0},\bm{I}/\alpha} \text{,} \;\;\;\;\;\;\; \vpost = \mathcal{N}\br{\theta|\bm{\mu},\text{diag}\br{\bm{\sigma}^2}}\text{,}
\end{equation*}
where ${\alpha>0}$ is a known precision parameter and $\bm{\mu} \in \mathbb{R}^{P}$, $\bms^2 \in \mathbb{R}^{P}$. $P$ is the number of the parameters in the network, i.e., the number of parameters in $\bmt$.
%where ${\alpha>0}$ is a known precision parameter and $\bm{\mu} \in \mathbb{R}^{\frac{P}{2}}$, $\bms \in \mathbb{R}^{\frac{P}{2}}$ are the mean and standard deviation of $\vpost$.
%$P$ is the number of parameters in the network, that is equal to double the number of dimensions in $\bmt$. 
$\vpost$ implies a factorization of the join in the product of the marginals, known as mean-field approximation. In VI, the variational parameters $\br{\bm{\mu},\bms^2}$ are obtained by maximizing the following objective:
\begin{equation}\label{eq:variational_objective}
    \mathcal{L}\br{\bm{\mu},\bm{\sigma}^2} = \sum_{i=1}^N \mathbb{E}_q\left[\log p\br{\Dcal|\bmt} \right] + \mathbb{E}_q\left[\log \frac{\prior}{\vpost} \right]\text{.}
\end{equation}
Eq. \eqref{eq:variational_objective} can be maximized with the gradient-based optimization, that is, with the following update:
\begin{equation} \label{eq:SGD_update}
       \bmm_{t+1} = \bmm_t + \rho_t \grad_{\bmm} \Lcal_t \;\;\;\;\;\textrm{and}\;\;\;\;\; \bms_{t+1} = \bms_t + \delta_t \grad_{\bms} \Lcal_t\text{,}
\end{equation}
where $t$ is the iteration index, $\grad_{x} \Lcal_t$ denotes an unbiased estimate of the gradient of $\Lcal$ at $(\bmm_t,\bms^2_t)$ with respect to $x$, and $\rho_t,\delta_t$ are adaptable learning rates.
The natural gradient VI method of \cite{khan_2017_conjugate} tackles the update \eqref{eq:SGD_update} in terms of the natural parameter $\bm{\alpha}$ of $\vpost$, rather than its mean and covariance matrix, and scales the gradient of the corresponding SGD update for $\bm{\alpha}_t$ with the inverse of the Fisher information matrix (FIM) of $\vpost$.
\cite{khan_2017_conjugate} show that the direct computation of the FIM can be avoided by computing {\it natural} gradients in the natural parameter space using the gradient with respect to the expectation parameters of the exponential-family posterior. For the Gaussian mean-field VI under consideration, this leads to the natural-gradient variational inference (NGVI) update:
\begin{align}
    \bmm_{t+1} &= \bmm_t + \beta_t \bms^2_{t+1} \odot \grad_{\bmm} \Lcal_t \text{,}\\
    \bms^{-2}_{t+1} &= \bms^{-2}_t -2 \beta_t \grad_{\bms^2} \Lcal_t \text{,}
\end{align}
with ${\beta_t>0}$ being a scalar learning rate. By expressing \eqref{eq:variational_objective} in terms of the standard MLE objective $f\br{\bmt} = -1/N\sum_{i=1}^N \log p\br{\Dcal_i|\bmt}$ and expressing the gradients of its expectation with respect to $\bmm$ and $\bms$ in terms of the gradient $g\br{\bmt}$ and Hessian $\mathbf{H}\br{\bmt}$ of $f\br{\bmt}$, 
%for the mean-field variant 
the update results in:
\begin{align}
    \bmm_{t+1} &= \bmm_t - \beta_t \br{g\br{\bmt_t}+\tilde{\alpha}\bmm_t} / \br{\bmss_{t+1}+\tilde{\alpha}} \text{,} \label{eq:VOGN_update_mu}\\
    \bmss_{t+1} &= \br{1-\beta_t}\bmss_t + \beta_t \text{diag}\br{\mathbf{H}\br{\bmt_t}} \text{,} \nonumber
\end{align}
where the division is element-wise, $\tilde{\alpha} {=} \alpha/N$ and $\bmt_t {\sim} \mathcal{N}\br{\bmt|\bmm_t,\text{diag}\br{\bms^2_t}}$, with $\bms^2_t {=} [{N(\bmss_t +\tilde{\alpha})}]^{-1}$.
The scaling vector $\bmss_t$ involves the gradient and the diagonal of the Hessian, which can be replaced by their stochastic estimates $\hat{g}\br{\bmt}$ and $\grad^2_{\bmt \bmt} f\br{\bmt}$. The former can be computed using backpropagation. Due to the general non-convexity of $f$, the latter can be negative, which might lead to negative variances. Non-negativity is granted by the following approximation:
\begin{equation} \label{eq:VOGN_gradient_approx}
    \grad^2_{\theta_j \theta_j} f\br{\bmt} \approx \frac{1}{M} \sum_{i\in \mathcal{M}} \left[\grad_{\theta_j} f_i\br{\theta}  \right]^2 := \hat{h}_j \br{\bmt} \text{,}
\end{equation}
with $i$ denoting the $i$-th data sample in the mini-batch $\mathcal{M}$ of size $M$, and $\theta_j$ the $j$-th element of $\bmt$.
By writing for  $\hat{\mathbf{h}}\br{\bmt_t}$ the vector of all $\hat{h}_j$, under this approximation the update for $\bmss_t$ reads:
\begin{equation}
    \bmss_{t+1} = \br{1-\beta_t}\bmss_t + \beta_t \hat{\mathbf{h}}\br{\bm_t} \text{.} \label{eq:VOGN_update_sigma}
\end{equation}
The algorithm involving updates \eqref{eq:VOGN_update_mu} and \eqref{eq:VOGN_update_sigma} is referred to as the Variational Online Gauss-Newton (VOGN) \cite{khan_2018_fast}. Opposed to SGD and related algorithms such as RMSprop, Adam, and AdaGrad, which use the gradient magnitude $[\frac{1}{M} \textstyle\sum_{i\in \mathcal{M}}\grad_{\theta_j} f_i\br{\theta}  ]^2$ for approximating the $j$-th entry of the diagonal Hessian, in \eqref{eq:VOGN_gradient_approx}, VOGN uses averages of the squared gradients, avoiding explicit constraints on $\bms^2$. 
As shown in \cite{osawa_practical_2019}, it leads to good empirical performance and practical feasibility of the updates \eqref{eq:VOGN_update_mu} and \eqref{eq:VOGN_update_sigma} on large datasets compared to alternatives, e.g. Bayes by Backprop (BBB)
\cite{blundell2015weight}. 

Existing automatic-differentiation libraries can be used to retrieve the gradients; however, current codebases directly return sums of the gradients over mini-batches, whereas individual gradients are required in \eqref{eq:VOGN_gradient_approx}. Thus via chain rule, we derive the individual gradients for a TABL layer and adapt current second-order optimization routines \cite{osawa_2018_torch} to accommodate them.
Our B-TABL implementation adopts a log-softmax activation function at the last layer such that the output vector of the network interprets as logs of class-probabilities, with a one-to-one mapping between classes and indexes of vectors' elements. That is, for an input time-series $\mathbf{X}_i$, the corresponding output of the network is a vector
\begin{equation*}
    \log \bm{p}_i = [\log p_{i1}, \log p_{i2}, \dots, \log p_{iC}],
\end{equation*}
where $p_{ic} =\exp{\br{\bm{l}_i[c]}}/{\sum_c \exp\br{\bm{l}_i[c]}}$, $c$ being an index running along the number of network outputs (the number of classes $C$). $\bm{l}_i$ denotes the output of the last layer, corresponding to the input $\mathbf{X}_i$, passed to the softmax activation, and $\bm{l}_i[c]$ its $c$-th element. The loss used is the negative log-likelihood, i.e., for a sample in class $c$ the loss is computed as $-\log\bm{p}_i[c]$. We shall refer to 
%$\log \bm{p}_i$ as log-score, and to
$\bm{p}_i$ as class-probabilities or scores. Losses are averaged across samples for each mini-batch. 
For a trained model and for each input vector, predictions are provided by the class of the maximum log-score (or maximum class-probability), i.e., $\max_{c} \log \bm{p}_i[c]$. We shall refer to this criterion as the classification rule.
Therefore, due to posterior sampling, log-scores are stochastic (as for the class-probabilities and the index of their maximum element), leading to stochastic class labels.

\section{Experiments}
\subsection{High-Frequency Limit Order Book Data}
Trading in modern financial markets is organized through an order-driven mechanism that collects and matches inflowing limit and market orders through a time-priority rule. Trades participate in the market by submitting orders or cancellations over previously submitted orders. Each message (order or cancellation) submitted to the exchange comes with an associated timestamp, price, and quantity (along with a unique identifier).
By submitting a limit order, a trader expresses his/her willingness to buy or sell a certain amount of the security at a specified price, i.e., the trader specifies the buy/sell price and the number (or fractions) of stocks he/she wants to trade. Limit orders are collected and stored in what is known as the Limit Order Book (LOB). At a time instance $t$, the cross-section of the LOB provides a snapshot of the number of outstanding limit orders, their prices, and quantities. 

In particular, buy (sell) limit orders define the bid (ask) side of the book. The highest buy and lowest ask prices represent the best prices to sell or buy a certain amount of a security. These best-prices are known respectively as bid and ask prices ($p^B_t, p^A_t$).
Market orders are immediately executed on the bid or ask side at the current best price, leading to trades.
Limit orders at the current bid/ask prices are filled according to a time-priority rule (first submitted, first traded). A market order decreases the quantity available at the best price, and, if the market order quantity is equal to or greater than the outstanding quantity of the limit orders at the current best price, it reduces the total depth of the market (i.e., the number of different price levels on which the limit orders are arranged).
As the limit orders on the top of the book are filled, the actual best price moves to that of the next LOB level until a new incoming limit order (on the same side of the book) re-fills the gap between the bid and ask prices, or a new market order erodes the top of the book causing a further update in the best bid or ask price. We refer to, e.g., \cite{ntakaris_benchmark_2018} for further details on the LOB mechanism.

It is clear that the order inflow (along with order cancellations) is governed by a highly-stochastic mechanism that leads to a rich, multidimensional dataset consisting of order types, prices, and quantities, whose instances reflect the dynamics of the bid and ask prices as well as of deeper LOB levels. Although broad stylized facts on the LOB dynamics lead to some analytic tools for modeling the LOB, e.g., \cite{cont_stochastic_2010}, tackling its dynamics is very challenging, and ML methods can provide a useful alternative for a number of forecasting goals. While the first level of the LOB has been commonly used in econometric research, \cite{tran2021levels} showed that the information in multiple levels increases the performance of ML models. Both ML and econometrics research focused on the dynamics of the synthetic price measure across the two sides of the book known as mid-price: $p_t = {\small \frac{1}{2}}  (p^A_t-p^B_t)$.

We focus on the task of forecasting mid-price changes at the future (tick-by-tick) updates of the LOB. This implies a complex classification problem over three classes: mid-price increases, mid-price decreases, or remains stationary. We used the publicly available FI-2010 dataset \cite{ntakaris_benchmark_2018}, which collects the LOB states for five stocks traded at the NASDAQ Nordic Helsinki exchange from June 1 to June 14, 2010 (collecting approximately 4.5 million events across ten trading days). At each epoch (i.e., LOB update), the data consists of 144-dimensional feature vectors. In total, there are 453,975 features extracted over non-overlapping blocks of ten events and normalized using the $z$-score. The dataset provides labels corresponding to the direction of the price movement on five different horizons, corresponding to the price movements in the next 10, 20, 30, 50, and 100 events. In our experiments, we utilize the 10-events horizon and adopt the experimental setup of \cite{tsantekidis_forecasting_2017,tsantekidis_using_2020} where the last three days are taken as the test set (corresponding to 150,418 samples). For the first seven days, the initial 75\% of instances constitute the training set, and the last 15\% the validation set.

\subsection{Experiment setting}

As in \cite{tran_temporal_2019}, we use the first 40 dimensions consisting of raw prices and quantities. 
For the BNN implementation, we employ the VOGN algorithm. Networks' weights are initialized under a multivariate Gaussian prior with parameters $\bm{\mu} \eq \bm{0}$ and $\bm{\Sigma} \eq \bm{I}$. 
The learning rate, momentum factor, and decay rate of the L2 norm regularization are respectively set to 0.01, 0.999, and 0.85.
Its performance on the validation set is evaluated over 10 MC draws from the posterior at each epoch, while the predictive distribution for each input in the test set is approximated by the collection of $N_{s} = 50$ forecasts following $N_{s}$ feed-forward passes for $N_{s}$ independent samples from the variational posterior. Prior's means and variances, respectively set to one and zero, are initially warmed up following the method described in \cite{osawa_practical_2019}.

The Bayesian training of the network is evaluated with respect to two non-Bayesian alternatives: the ADAM \cite{kingma_adam_2015} optimizer and Stochastic Gradient Descent (SGD). For SGD, the momentum is set to 0.99; for ADAM, the first and second moments are fixed to 0.9 and 0.999. For both algorithms, the initial learning rate is set to 0.01 and dynamically updated until the validation loss reaches a plateau.
For all the optimizers, the training is set to 1000 epochs with a mini-batch of size 256.
When training with ADAM, we also employ MC dropout \cite{gal_dropout_2016} in the testing phase. 
MC dropout is not a Bayesian method but has a connection with Bayesian theory and serves as an approach to predictive distribution approximation \cite{gal_dropout_2016}. A random deletion of the NN connections allows for posterior sampling. Sampling from the approximate posterior enables MC integration of the likelihood, uncovering an approximation to the predictive distribution. 
By repeated forward passes for the same input sample, the randomized dropout yields samples from the predictive distribution. \cite{gal_dropout_2016} find that even a small number of forward passes can suffice. Similar to the B-TABL, we apply $N_{s} = 50$, and set the dropout rate to 10\%. As in \cite{osawa_practical_2019}, we do not compare VOGN with Bayes by Backprop \cite{blundell2015weight} since it is very slow to converge for larger-scale experiments like the one targeted in this paper.

\section{Results}
\subsection{Model calibration and learning curves}  % REVIEWED

\begin{figure}[htbp]
    \centering
    %\scalebox{0.99}{
    \includegraphics[trim={1.5cm 0.5cm 1.5cm 1.2cm},clip,width=\columnwidth]{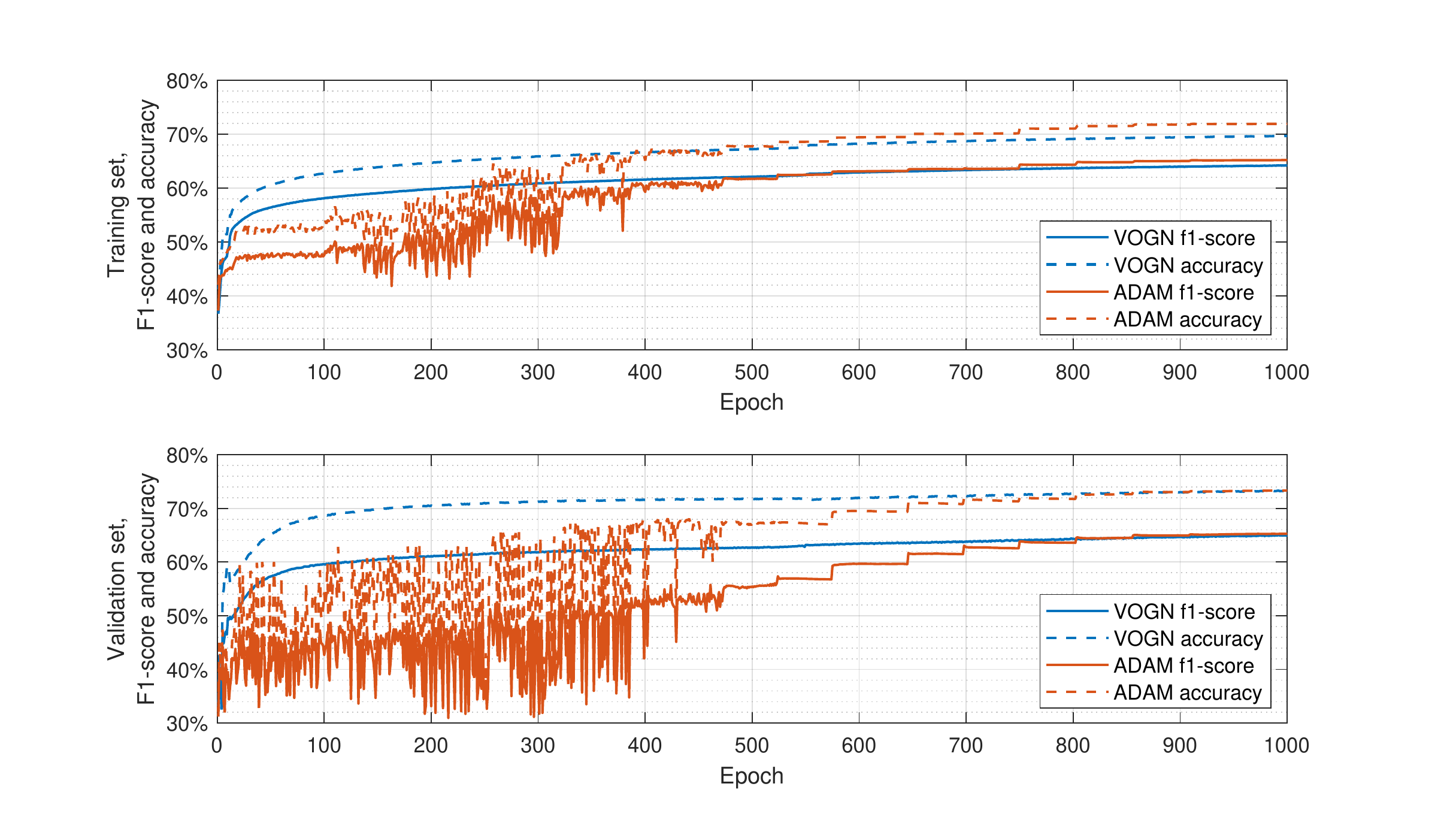}
    %}
    \caption{Learning curves for VOGN and ADAM for the training set (upper panel) and validation set (lower panel).}
    \label{fig:learning}
\end{figure}

In \figurename ~\ref{fig:learning} we compare f1-scores and accuracy metrics across training epochs for both training and validation sets. For both VOGN and ADAM, it takes as little as 15 epochs to stabilize and smooth the learning rate of the curves. The initial values of the parameters are randomly initialized. To avoid local minima and to boost the search, the learning rate in ADAM is regularly perturbed, resulting in the step-wise behavior observed of the curves observed across the panels. 

For VOGN, curves referring to the training set show a steeper rate at initial epochs and up to about epoch 500, reporting a remarkably higher f1-score and accuracy than for ADAM. This stands for a general superiority in the performance of VOGN with respect to ADAM at early epochs up to moderate ones, indicating that despite the random initialization and the stochastic components embedded in VOGN, perhaps due to its higher number of parameters due to the existence of the parameters' variance, the algorithm converges rather quickly. Only around epoch 500 ADAM metrics are comparable to those of VOGN. At higher epochs, we do not observe a relevant difference in f1-scores, while, in terms of accuracy, ADAM slightly outperforms VOGN on the training set. That is, the learning in ADAM is shown to be slower but, on average, steady, with certainly lower rates compared to VOGN, but constantly improving across the epochs. On the other hand, after steep improvements in initial phases, VOGN's training is quite achieved already at epoch 500, leaving only a slight 5\% improvement of the metrics in the following 500 epochs.

Also on the validation set, we observe that at initial epochs, metrics for VOGN greatly outperform those for ADAM; perhaps the prior variance adds a randomization effect that allows a wider sampling of the space around the local parameters to access large gradients that readily adjust the step direction towards the minimum. It is only around epoch 800 that we observe a comparable performance. This could be interpreted as a better generalization ability of VOGN on unseen data, especially if noticing that for VOGN the f1-score and accuracy curves on the validation set are slightly higher than for training. The homogeneity of the data and the same complexity across the two sets, further motivate the conclusion that VOGN embeds a more general classification rule, along with the metrics provided in \ref{app:stock_perf}. Also the different levels in curves' smoothness underline that while VOGN quickly approaches the minimization objective, ADAM appears to repeatedly overshoot the objective, leading to segmented curves up to epoch 500.
At higher training epochs, validation curves' rates of growth for VOGN and ADAM appear quite flat, indicating that the training is overflown and the performance metrics are comparable. In this light, we might expect a comparable performance of the two algorithms on the test set, perhaps without a strong winner. This is indeed the case, see Subsection \ref{sec:perf_measures} and Subsection \ref{app:other_analyses}. As it also emerges from Subsection \ref{sec:perf_measures}, the performances for MCD and SGD are quite poor compared to VOGN and ADAM, making the former two optimizers quite unsuitable for our classification task, and therefore omitted from \figurename ~\ref{fig:learning}.

\subsection{Posterior distribution}  % REVIEWED

\begin{figure}[htbp]
    \centering
    %\scalebox{0.99}{
    \includegraphics[trim={1cm 0.35cm 0.3cm 0.5cm},clip,width=\columnwidth]{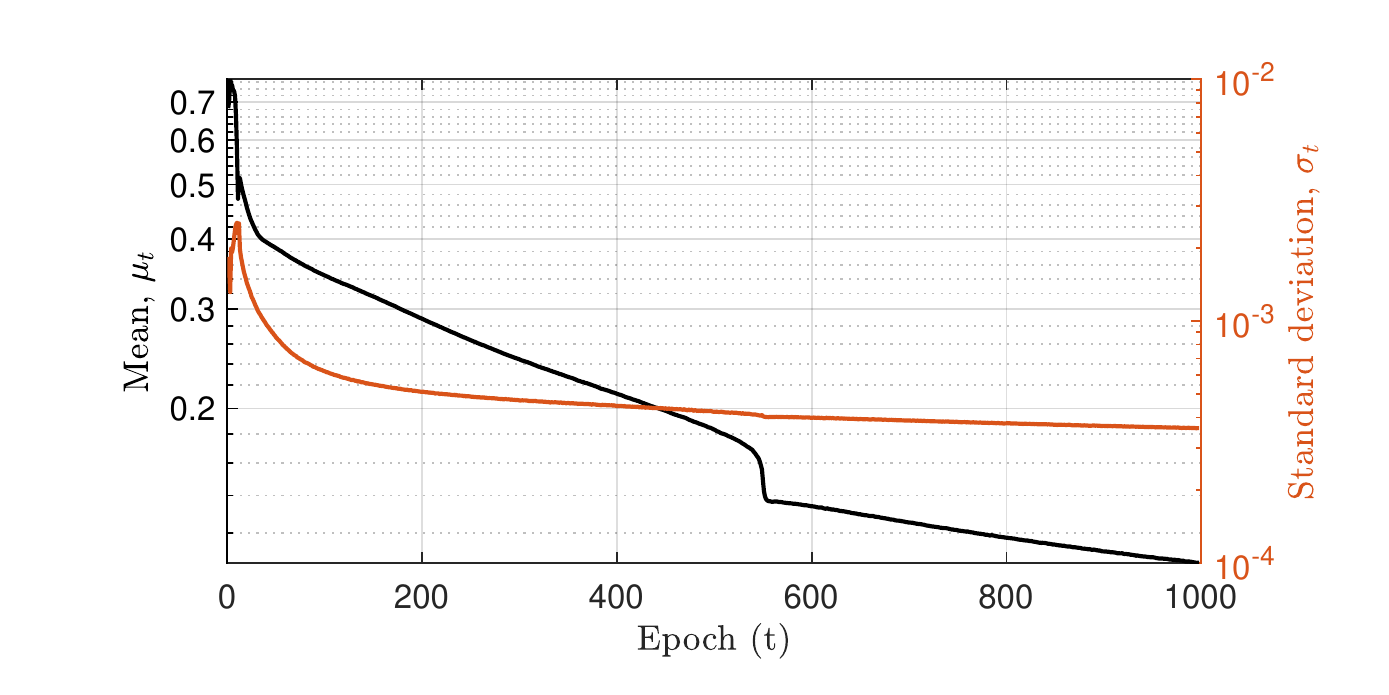}
    %}
    \caption{Learning of the variational parameters for TABL's mixing coefficient $\lambda$.}
    \label{fig:posterior}
\end{figure}

Following the updates \eqref{eq:VOGN_update_mu} and \eqref{eq:VOGN_update_sigma}, VOGN learns variational posterior's mean and updates the prior precision to the posterior diagonal covariance matrix $\bm{\sigma}^2 I$, with $\bm{\sigma}^2_t = 1/\br{N\br{\bm{s}_t+\tilde{\alpha}}}$. For illustration purposes, \figurename ~\ref{fig:posterior} depicts the learning of B-TABL's $\lambda$, a characterizing parameter for the network architecture.

In general, for all the parameters in TABL we observe a similar pattern where both parameters' means and variances converge to certain (different) levels. The posterior distribution is representative of the parameters' uncertainty after observing the data, that in VOGN's variational setting is forced to be a Gaussian distribution. We make use of the posterior's valuable information on the parameters' relevance by conducting individual t-tests on their significance. Very low p-values support the relevance of all the TABL parameters and implicitly that the network architecture is well-scaled for the problem under consideration. Following (\ref{eq:predictive}), the posterior distribution builds the predictive one, a major focus in this paper.

\subsection{Predictive distribution}\label{sec:predictive} 

\subsubsection{Interpreting predictive probabilities}\label{subsec:interpetation_predictive}  % REVIEWED
\begin{figure*}[h]
    \centering
    \scalebox{0.8}{
    \includegraphics[trim={2.5cm 0.5cm 2.5cm 1cm},clip,width=2\columnwidth]{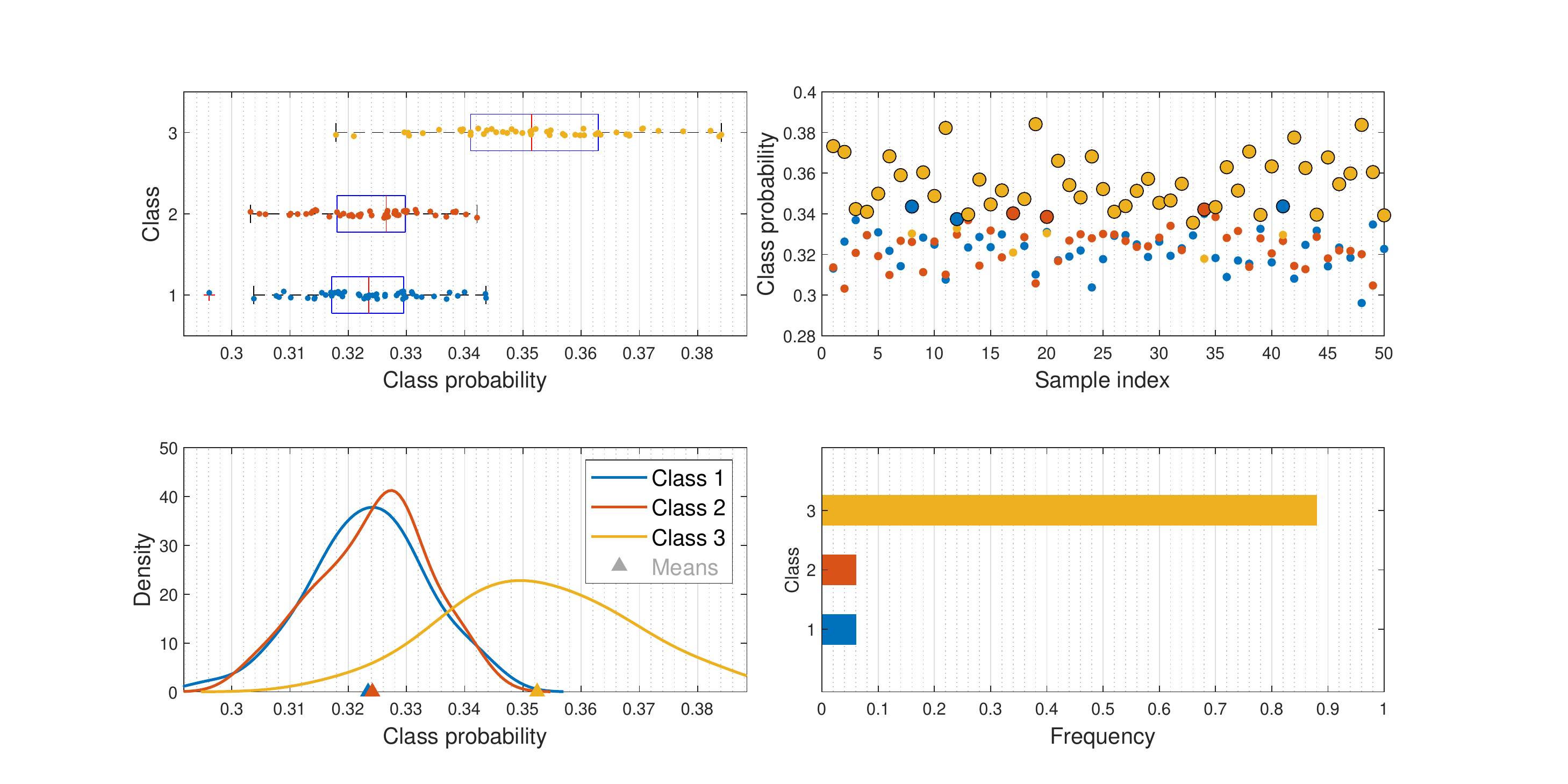}
    }
    \caption{Class-probabilities and forecasts for a typical test example. Top-left, panel (1): box-plots of class-probabilities. Bottom-left, panel (2): kernel density estimates and means of class-probabilities. Top-right, panel (3): class-probabilities per class, highlighting those of maximum probability. Bottom-right, panel (4): histogram of forecasts' labels.}
    \label{fig:boxplot}
\end{figure*}

We approximate VOGN's predictive distribution with $N_s=50$ draws from the posterior distribution. I.e., according to \eqref{eq:bayesian_model_averaging}, we approximate the posterior distribution in \eqref{eq:predictive} for a given input sample by fifty samples drawn from it to capture the uncertainly associated with a forecast $\hat{y}_{ic}$ given an unseen input $\bmx_i$ and the data used in the training phase. According to the decision criterion in \eqref{eq:classification_crit}, the forecast's class is given by the predicted class of maximum class-probability. Note that, aligned with \eqref{eq:predictive}, the predictive distribution is a distribution on class-probabilities $p\br{\bmy_i|\bmx_i,\Dcal}$ and not on the forecasts $\hat{y}_{ic}$. \figurename ~\ref{fig:boxplot} provides insights into interpreting predictive probabilities and explains pitfalls in uncertainty interpretation. 

The bottom-right panel in Figure~\ref{fig:boxplot} depicts the box-plot of the class-probabilities corresponding to $N_s$ forward passes for a certain input in the test set. For this example, the distribution of the predicted classes is unarguably leaning towards class 3, while across the 50 samples, the class probabilities for classes 1 and 2 are similar. The bottom-left panel depicts the corresponding kernel probability density functions of the class probabilities, along with their per-class average. I.e., predicted probabilities for the sample input in \figurename ~\ref{fig:boxplot} are $p\br{\bmy_i|\bmx_i,\Dcal} = \left[0.323,0.324,0.352\right]$  for classes 1, 2 and 3  respectively. 
A complete analysis of the joint distribution of $p\br{\bmy_i|\bmx_i,\Dcal}$, here out of scope as it would likely involve a dependence modeling step through copulas that perhaps is not general. Yet the extracted mean values are quite informative. In particular, we see that the probability of class 3 is certainly higher than those of other classes, even though not overwhelming.
That is, the predictive distribution underlines a scenario of rather high uncertainty. Indeed, an investor who, e.g., invests in an asset expecting its price to increase would, on average, observe an actual increase in the asset value with a 35\% probability. This is the correct interpretation upon the predictive probability  $p\br{\bmy_i|\bmx_i,\Dcal}$. However, upon applying the classification criterion \eqref{eq:classification_crit} the conclusions might be quite misleading. The top-right panel in \figurename ~\ref{fig:boxplot} depicts per-sample class-probabilities, where big-sized points represent classes of maximum probability. For 43 samples out of 50 the class of maximum probability is 3. For three samples is class 1, and for three samples is class 2. By applying \eqref{eq:classification_crit} we would classify 86\% of the samples in class 3, as depicted by the histogram in panel 4 of \figurename ~\ref{fig:boxplot}. The histogram misleadingly covers a situation of high uncertainty with a quite overwhelming frequency observed for class 3: if we were to draw the forecasts according to the joint distribution in panel 2 we would observe, on average, only about 35\% of the samples in class 3.

The availability of the predictive distribution in Bayesian DL frameworks is certainly the most remarkable aspect with respect to non-Bayesian approaches such as ADAM. Indeed, continuing with the above example, ADAM would not capture any uncertainty in the predicted label. An investor following ADAM's forecasts (whose performance is comparable with that of VOGN, see Section \ref{sec:perf_measures}) would not be able to capture the high degree of uncertainty that VOGN unveils. Needless to say that the impact of uncertainties on whatever trading strategy an investor adopts is quite significant. E.g., an investor might choose to trade based only on predictions associated with relatively low uncertainty, or take well-informed actions to account for the actual possibility that the direction of the price movement is opposite to the predicted one. On the other hand, ADAM's forecasts are incapable of addressing the low 35\% probability chances of a price increase, leaving the investor completely blind about the actual probability of a price increase and the perhaps adverse downward 32\%-likely movement.

By inspecting a large number of examples and the overall statistics on class-probabilities we observe that the case studied in \figurename ~\ref{fig:boxplot} is somewhat atypical, in the sense that ~35\% of the predicted probabilities for the maximum-probability class are in the lowest quantiles of predicted probabilities for the prediction's class, see \ref{subsec:predictive}. Typical values are about 50\%: this still results in a wildly uncertain general scenario. Such relevant uncertainty for operational scenarios and real-life decision processes is entirely left unaddressed by ADAM and non-Bayesian methods. Whereas the decision criterion \eqref{eq:classification_crit} does provide a feasible and practical way to construct forecasts, average class-probabilities on repeated forward passes (i.e., predictive distributions) are the truly informative element about forecasts' uncertainty, inaccessible to non-Bayesian DL approaches.

\subsubsection{Predictive probability for the maximum-probability class}\label{subsec:predictive} % REVIEWED
 
or the sample input in \figurename ~\ref{fig:boxplot} the true label corresponds to class 1. In general, for wrong and correct classifications, the score variation in the class of maximum-probability class can be large. 
Table \ref{tab:predictive} reports some statistics on the class of maximum probability (rank 1, denoted by for an input $\bmx_i$ with $\hat{p}^{(1)}_i$ ), class of second-highest maximum probability (rank 2, $\hat{p}^{(2)}_i$) and on the remaining class (rank 3, $\hat{p}^{(3)}_i$), along with the difference between the first two ranks, for correctly and miss-classified labels. 
For correct classifications, the average (median) predictive probability on rank-1 classes is 55\% (51\%), while for the miss-classified ones 51\% (50\%), the average distance between the predictive probabilities on classes of rank 1 and 2 is respectively 30\% (25\%) for correct classifications and 22\% (20\%) for miss-classifications. 
In both cases, we do observe samples of probability rank 1 with a corresponding predictive probability as high as 100\% and as low as 33\%.
Ideally, we would like to observe (i) high rank-1 predictive probabilities for correct classifications, (ii) quite lower values for miss-classified samples, and (iii) a neat separation between $\hat{p}^{(1)}_i$ and $\hat{p}^{(2)}_i$ for correctly classified samples. This is the case, but the magnitudes of the differences are small. 
Table \ref{tab:predictive} underlines that the predictive uncertainty in the forecasts is consistent and homogeneous whether the labels are eventually correct or wrong, with $\hat{p}^{(1)}_i$ consistently being of about twice $\hat{p}^{(2)}_i$ and $\hat{p}^{(2)}_i$. The observed differences in rank-1 and rank-2 predictive probabilities are, on average, as little as 7.5\% between correctly and miss-correctly classified labels, while $\hat{p}^{(1)}_i$ differs by only 3.4\%.

\begin{figure}[htbp]
\centering
\scalebox{0.8}{ 
\includegraphics[trim={0.5cm 0.5cm 0.5cm 0.5cm},clip]{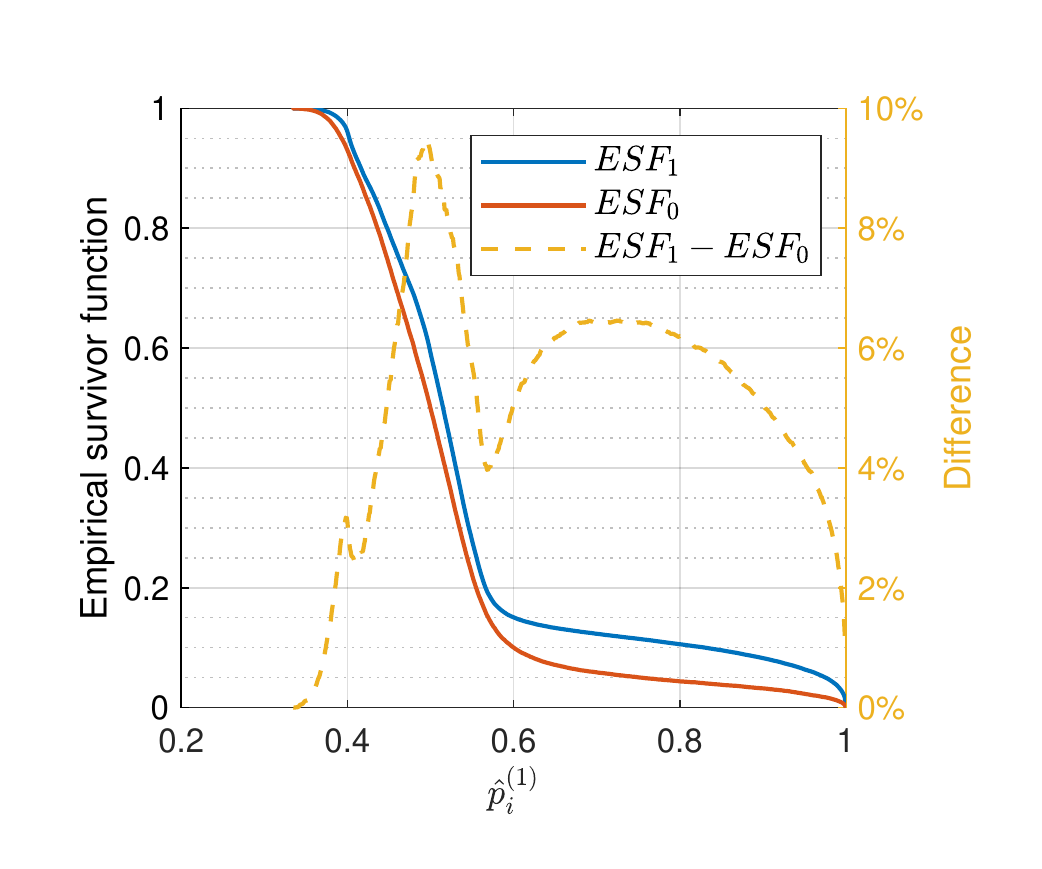}
}
\caption{Empirical survivor function of $\hat{p}^{(1)}_i$ for correct classifications ($ESF_1$) and miss-correct ($ESF_0$), along with their difference.}
\label{fig:ESF}
\end{figure}

Accordingly, the empirical survivor function (ESF) for correctly classified labels in \figurename ~\ref{fig:ESF} slightly dominates the one for miss-classified labels. Importantly, \figurename ~\ref{fig:ESF} unveils that the probability of observing $\hat{p}^{(1)}_i> 0.6$ is only about 10-15\%. This means that mild-to-low uncertainties on the maximum probability class are quite rare, and high-confidence is even rarer (7\%-9\% for $\hat{p}^{(1)}_i> 0.9$). However, the ESFs do not cross each other, and the difference is positive. For the same (or greater) level of confidence, the number of correctly classified samples is, on average, 5\%  (but up to 9\%) higher for the correctly classified samples than the miss-classified ones (difference curve in \figurename ~\ref{fig:ESF}).

\begin{table}[htbp]
\caption{Statistics on VOGN's predictive probabilities.}
    \centering
\begin{tabular}{lcccc}
\toprule
      & \multicolumn{3}{c}{Class-probability} & Difference \\
& $\hat{p}^{(1)}_i$ & $\hat{p}^{(2)}_i$ & $\hat{p}^{(3)}_i$ & $\hat{p}^{(1)}_i-\hat{p}^{(2)}_i$ \\[0ex]
\cmidrule(lr){2-4} \cmidrule(lr){5-5}     \multicolumn{5}{l}{\textit{\textbf{Correctly classified labels}}} \\
Mean  & 0.550 & 0.254 & 0.196 & 0.296 \\
Median & 0.515 & 0.262 & 0.220 & 0.250 \\
Min   & 0.335 & 0.000 & 0.000 & 0.000 \\
Max   & 1.000 & 0.478 & 0.330 & 1.000 \\[1ex]
\multicolumn{5}{l}{\textit{\textbf{Miss-classified labels}}} \\
Mean  & 0.516 & 0.295 & 0.189 & 0.220 \\
Median & 0.500 & 0.296 & 0.204 & 0.196 \\
Min   & 0.335 & 0.000 & 0.000 & 0.000 \\
Max   & 1.000 & 0.479 & 0.330 & 1.000 \\
\bottomrule
\end{tabular}
    \label{tab:predictive}
\end{table}

\subsubsection{Distribution of the scores} % REVIEWED

\begin{figure*}[htbp]
    \centering
    \scalebox{0.85}{
    \includegraphics[trim={1.5cm 0.3cm 1.5cm 0.5cm},clip,width=2\columnwidth]{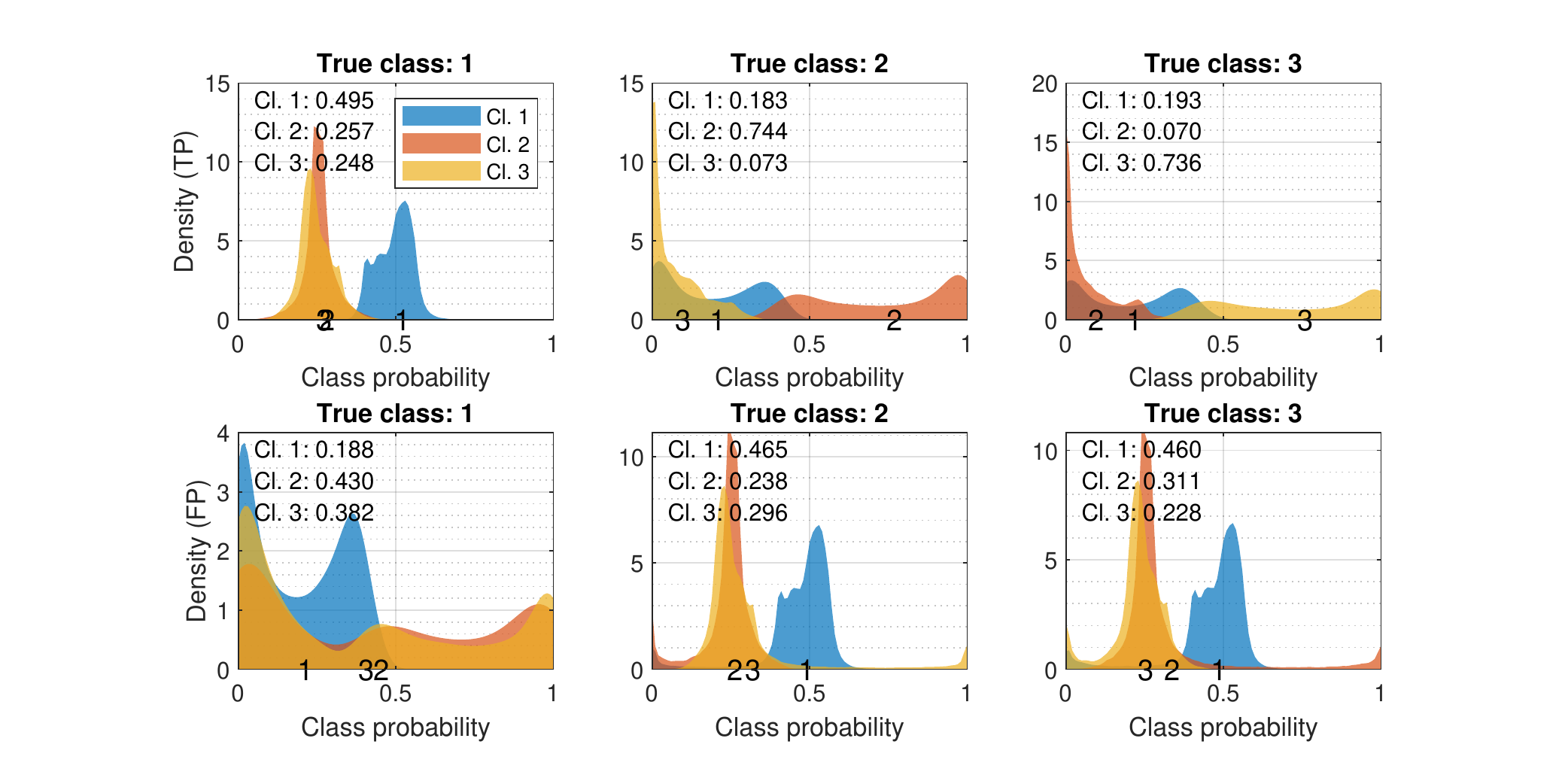}
    }
    \caption{Distribution of VOGN's predictive probabilities. Top row: distribution of the class-probabilities for correctly-classified labels. Bottom row: distribution of the class-probabilities for miss-classified labels. Labels on the x-axis are in correspondence of class-probability averages, also overprinted in the top-left corners.}
    \label{fig:scores_density}
\end{figure*}

To better understand the uncertainties arising from the predictive distribution, \figurename ~\ref{fig:scores_density} depicts the (kernel) density estimates of the scores across correctly and miss-classified labels. The top row in \figurename ~\ref{fig:scores_density} refers to correctly classified samples (TPs). For TPs, scores are well-separated in the sense that the distribution of class one is well-detached and distinguishable from the others.
Interestingly, when the model is correct, the uncertainty in classes 2 and 3 is much lower than in the stationary-price case. This stands for the existence of clear patterns in the features that are truly indicative of the direction of the price movement, driving predictive probabilities close to 1 (i.e., uncertainty close to zero). When the model is correct about class-1 assignments, its confidence is somewhat lower and the densities of the scores for whatever change in price direction generally overlap. Confidence in stationary prices is about 0.5, while the remaining 0.5 is equally spread across classes 2 and 3. 

Indeed, by only focusing on the miss-classified labels (FPs) in the bottom row in \figurename ~\ref{fig:scores_density}, we find further evidence that when the model does not correctly classify a stationary mid-price, its predictions are about equally spread among a price-increase and a price-decrease, showing that in this case there is no intrinsic bias in the model parameters leaning towards a certain class; the model is simply wrong, and forecasts are flip-coins on classes 2 and 3. On the other hand, the bias towards the majority class is consistent for FPs in classes 2 or 3, and the scores for the true-label are always those of lowest means. 
The same distribution on class-1 TPs almost identically replicates on class 2 and 3 FPs: the model interprets certain patterns in the features as remarkably non-indicative of the true class 2 and 3 labels, causing an overflow of low scores for both of them. The relevant probability mass, excluded from classes 2 and 3, is transferred to class 1 following a distribution being very close to that observed on class-1 TPs. This suggests that the model well-distinguishes patterns indicative of classes 2 and 3, and when these are absent, class-1 classification is enforced. In this regard, see \ref{subsubsec:model_learning}.

Tails in FPs for classes 2 and 3 constitute interesting cases of very high class-3 and class-2 predictive probabilities corresponding to wrong assignments in classes 2 and 3. Patterns indicative of classes 3 and 2 are causing false positives in classes 2 and 3: (rarely) typical features for classes 3 and 2 are observed for mid-prices, eventually moving in the opposite direction. These real-surprise elements in the order flow are perhaps aligned with its stochastic nature.

\subsubsection{Model learning}\label{subsubsec:model_learning} % REVIEWED

There is a conclusive important insight on the model's learning within figure \figurename ~\ref{fig:scores_density}. In particular, the B-TABL (and TABL, as the following applies to the scores from all the optimizers) architecture is learning how to classify increases and decreases in mid-prices but not stationary prices. The distribution of class-1 scores is the very same for the TPs in class 1 and the FPs in classes 2 and 3. That is, a distribution is placed over the scores in class 1 and is updated only if relevant features for class-2 or class-3 decisions are detected. Indeed when the model is not capable of correctly classifying classes 2 and 3, the distribution of the scores on classes 2 and 3 is roughly the same. In fact, the three plots in \figurename ~\ref{fig:scores_density} are almost identical at a visual inspection. I.e., unless there is robust evidence of an upward or downward price movement very likely corresponding to correct classification, the distribution of class-1 scores as FPs for classes 2 and 3 is the same. When the model truly does not detect any relevant information to discern whether the price is moving, movements are classified as stationary, and the very same distribution observed in class-1 TPs is adopted. This means that the density for TPs in class 1 is not actually learned from features that characterize this class. The density that is observed for class-1 labels is to be interpreted as the one that best improves the validation loss when the model is not detecting clear signals of future price movements. This is further supported by the analyses in Section \ref{subsec:labels_forecasts}, \ref{subsec:ROCandCC} and \ref{app:other_analyses}.

\subsection{Labels' forecasts} \label{subsec:labels_forecasts} % REVIEWED

Non-Bayesian methods such as ADAM and SGD provide single forecasts for a trained model. VOGN and MCD, due to the random sampling from the posterior and the dropout layer, lead to different foretasted labels at each forward pass. As noted in Section \ref{subsec:interpetation_predictive}, the distribution of the scores provides a misleading interpretation of forecasts' uncertainties. However, scores embed rich information for understanding the behavior of the labels' forecast and the underlining label-assignment mechanism. 
Again, the distribution of labels' forecasts analyzed in this section is based on $Ns=50$ draws.
For VOGN, 96\% of drawn labels are all assigned in the same class, 4\% to two classes, 0.3\% to three. Among the inputs whose labels' forecasts are across two classes, the predictive probability for the maximum class is 42\%, and the difference between the predictive probability on the two classes is, on average, 3\% (max. 36\%, min. 0\%). 45\% of these inputs correspond to true labels in class 1, and for 98\%, class 1 represents one of the two classes where the labels distribute, while the others are classes 2 and 3 with 50\% frequency. I.e., the model appears inconsistent in labeling stationary prices over positive or negative movements but is very consistent in labeling the latest two. 61\% of the samples with forecasts across two labels show, however, at least 80\% of the 50 draws in the same class, while very ambiguous inputs account for only 6\%, with a difference in the number of samples in the two classes not exceeding three. We could provide analogous information for the 0.3\% of the samples whose forecasts' labels are observed over three classes; rather, we point out that the predictive probability on the three classes is on average 31\%, 35\%, and 35\%, corresponding indeed to the most uncertain classifications. The above numbers suggest that typically the consistency in the foretasted labels is remarkable, i.e., that the modal value is overwhelming. This means that a single draw from the network would be very likely to equal the modal value. Indeed by randomly selecting with replacement 500 output vectors of labels from the 50 draws available for each test example, on average, 99.36\% of the labels correspond to theirs distribution modal value.

Therefore in the following analyses, we include performance metrics based on modal forecasts as representative of the typical performance observed over a single forward pass. For completeness, we also consider metrics based on means and medians of the $N_s$ labels, rounded to the closest class' integer index. We omit the above details for MCD but underline that 72\% of the labels' forecasts are observed over three classes: a remarkable difference. This is due to the fact that on a single TABL layer, the regularization usually provided by dropout causes a random-like assignment of the output classes.

\subsection{Performance measures} \label{sec:perf_measures} % REVIEWED

Table \ref{tab:perf_all} reports the performance of different Bayesian and non-Bayesian optimizers for the (B)-TABL architecture. 
With respect to the test set, Table~\ref{tab:perf_all} includes micro-, macro-, and weighted macro- averages as synthetic measures for evaluating the overall performance of the different classifiers across multiple classes. 
Micro-averages are constructed by summing the true/false-positive/negative rates individually for each class, before applying the definition of the specific performance measure under consideration. 
On the other hand, macro-averages refer to simple averages of the individual performance measures computed for each class. 
By accounting for the relative sample frequency of each class in taking averages, we construct weighted macro-averages.
Note that accuracy and micro-averages for precision, recall, and F1-score are all equal and reported under a single column.
Although macro-averages are the performance measures usually reported, as our sample is highly imbalanced (67\% of the test samples in the stationary class and equally distributed across the remaining two classes), alternative multi-class statistics are here relevant. Macro-averages weight each class equally by computing the average of the metrics computed independently for each class. As a consequence,
non-dominant-class' metrics might mislead the conclusions on the overall performance of the classifier.
By accounting for class weights, weighted macro-averages naturally alleviate this issue. On the other hand, micro-averages, by summing the true/false-positive/negative rates individually for each class, aggregate the contributions of all classes to compute average metrics. By weighting each sample equally, micro-averages apply well to imbalanced problems where, from a qualitative standpoint, there are no differences in the importance of each class. In our context of imbalanced classes and multi-class task, the preferred metrics are the f1-score, which embeds both precision and recall, and micro-averages.

\begin{table*}[htbp]
\caption{Performance measures for the multi-class classification task. Notes: sample-by-sample refers to metrics computed for each of the $N_s$ samples. $\hat{Y}_{pred}$ (med.) refers to forecasts obtained from the predictive distribution based on the sample median (instead of sample average \eqref{eq:bayesian_model_averaging}, as for $\hat{Y}_{pred}$). MCD entries refer sample-by-sample averages.  MCD (pred.) refers to forecasts based on the predictive distribution. Top metrics are reported in bold, excluding the row referring to the sample-by-sample maximum (Max row).}
    \centering
\begin{tabular}{lccccccc}
\toprule
      & Any   & \multicolumn{2}{c}{Precision} & \multicolumn{2}{c}{Recall} & \multicolumn{2}{c}{f1-score} \\
      & Micro & Macro & Weighted & Macro & Weighted & Macro & Weighted \\
\cmidrule(lr){2-2}\cmidrule(lr){3-4}\cmidrule(lr){5-6}\cmidrule(lr){7-8}\multicolumn{8}{l}{\textit{\textbf{VOGN sample-by-sample}}} \\
Mean  & 0.774 & 0.736 & 0.763 & 0.592 & 0.774 & 0.636 & 0.751 \\
Median & 0.774 & 0.736 & 0.763 & 0.592 & 0.774 & 0.636 & 0.751 \\
Min   & 0.772 & 0.730 & 0.761 & 0.589 & 0.772 & 0.633 & 0.749 \\
Max   & 0.776 & 0.743 & 0.766 & 0.596 & 0.776 & 0.638 & 0.752 \\[1ex]
\multicolumn{8}{l}{\textit{\textbf{VOGN based on forecasts' function}}} \\
$\text{Mean}(\hat{Y}_i)$ & 0.772 & 0.731 & 0.761 & 0.591 & 0.772 & 0.633 & 0.749 \\
$\text{Median}(\hat{Y}_i)$ & 0.774 & 0.736 & 0.763 & 0.592 & 0.774 & 0.636 & 0.751 \\
$\text{Mode}(\hat{Y}_i)$ & 0.774 & 0.737 & 0.763 & 0.592 & 0.774 & 0.636 & 0.751 \\[1ex]
\multicolumn{8}{l}{\textit{\textbf{VOGN predictive distribution}}} \\
$\hat{Y}_{pred}$ & \textbf{0.774} & 0.737 & 0.763 & \textbf{0.592} & \textbf{0.774} & \textbf{0.636} & \textbf{0.751} \\
$\hat{Y}_{pred}$ (med.) & 0.774 & 0.737 & 0.763 & 0.592 & 0.774 & 0.636 & 0.751 \\[1ex]
\multicolumn{8}{l}{\textit{\textbf{Other optimizers}}} \\
ADAM  & 0.772 & \textbf{0.767} & \textbf{0.770} & 0.570 & 0.772 & 0.619 & 0.741 \\
MCD (mea.)  & 0.581 & 0.450 & 0.598 & 0.460 & 0.581 & 0.454 & 0.588 \\
MCD (pred.) & 0.638 & 0.500 & 0.630 & 0.492 & 0.638 & 0.495 & 0.634 \\
SGD   & 0.687 & 0.556 & 0.660 & 0.505 & 0.687 & 0.522 & 0.667 \\
\midrule\\[-2ex]
\multicolumn{8}{l}{\textit{\textbf{Differences}}} \\
Min - ADAM & 0.0\% & -3.8\% & -0.9\% & 1.8\% & 0.0\% & 1.3\% & 0.7\% \\
$\hat{Y}_{pred}$ - ADAM & 0.2\% & -3.1\% & -0.7\% & 2.2\% & 0.2\% & 1.6\% & 0.9\% \\
$\hat{Y}_{pred}$ - MCD (pred.) & 13.6\% & 23.7\% & 13.3\% & 10.0\% & 13.6\% & 14.0\% & 11.7\% \\
$\hat{Y}_{pred}$ - SGD & 8.7\% & 18.1\% & 10.4\% & 8.7\% & 8.7\% & 11.4\% & 8.3\% \\
\bottomrule
\end{tabular}%
\label{tab:perf_all}
\end{table*}

For the VOGN optimizer, results are divided into three panels. The upper one reports summary statistics for individual metrics computed for each of the $N_s=50$ simulated outputs, i.e., in a sample-by-sample fashion. 
Following the discussion in Section \ref{subsec:labels_forecasts}, the second panel addresses the possibility of constructing labels' forecasts based on group statistics extracted from the $N_s$ labels. These correspond to forecasts' labels sample mean, median, and mode. The former statistics requires rounding to the nearest integer to be feasible, yet in our sample rounding applies to only 3.5\% of the per-example labels' means, to 0.26\% of medians, and never to modes.
The third panel reports the metrics corresponding to predictive distribution, i.e., by considering the class of maximum predictive probability, computed under the standard Bayesian averaging approach and, alternatively, by considering median probabilities as a robust alternative to possible severe outliers.

For VOGN, predictive's distribution results are consistently the highest ones. However, up to three decimals, there are generally no differences between the three panels.  
Performance measures for median and modal forecasts largely overlap and equal predictive's distribution metrics. Slightly worse results are obtained by considering (rounded) forecasts' averages. The former aligns with the sample-by-sample centrality measures and predictive distributions' ones. This also suggests that for forecasting purposes, a single draw from the posteriors (whose corresponding label would approximate the forecasts' median label very closely) would lead to results perfectly aligned with the predictive's ones (implying a considerable computational advantage).

Among the other optimizers, ADAM stands out as the most valid alternative. Expect on precision, it does not perform better than any VOGN's metrics. Interestingly, metrics' minima in the top panel are always higher than ADAM's metrics (except for precision, where neither the maximum reaches ADAM's performance). This provides significance to the results in favor of VOGN as even the most unfortunate posterior sampling shows superior performance than ADAM, up to 1.8\%.
Concerning VOGN's predictive distribution, the observed improvements in performance with respect to ADAM are slight yet significant: the Bayesian optimizer does not provide worse results than the widely-adopted ADAM (except for precision), and it enables the predictive analysis of forecasts' uncertainty described in Section \ref{sec:predictive}. Lastly, MCD and SGD do not seem to be competitive for the prediction task under analysis. In \ref{app:stock_perf} we provide analogous stock-specific results.

Our following considerations concern the single-class problem classification.
Despite the above multi-class task where each label is classified across three classes, by the single-class task we mean a binary classification problem where the true class is specified in advance, and the other classes constitute the negative class.
Remind that the model is calibrated for the multi-class task: single-class metrics could be improved by re-calibrating the model specifically for forecasting a specific price-change direction.

\begin{figure}[htbp]
    \centering
    %\scalebox{0.95}{
    \includegraphics[trim={0.5cm 0.3cm 0.5cm 0.5cm},clip,width=\columnwidth]{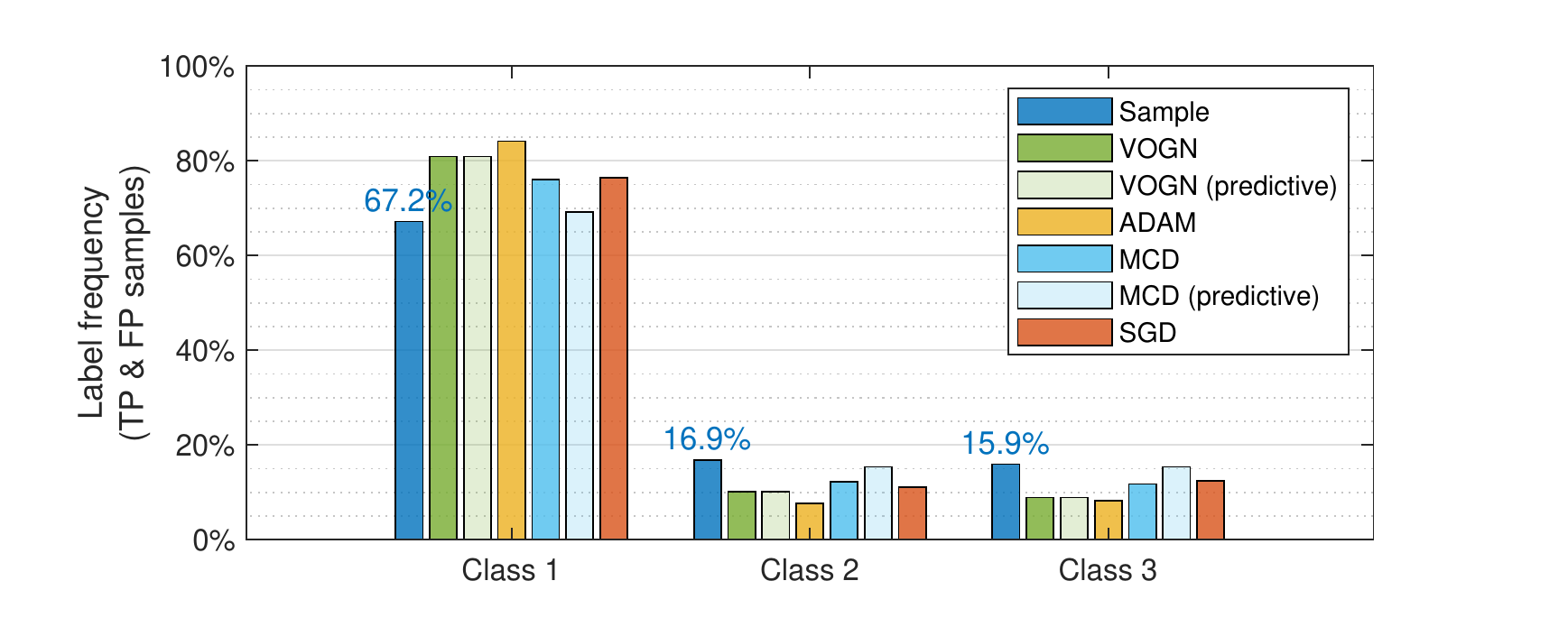}
    %}
    \caption{Overall distribution of labels' frequencies across the classes. The actual sample-data distribution is meant to be used as a benchmark. Bars display for each class and optimizer the fraction of labels correctly (TPs) and incorrectly (FPs) assigned to it.}
    \label{fig:Bar_all_cases}
\end{figure}

A first useful analysis is that of inspecting the distribution of labels assigned to the true class, see \figurename ~\ref{fig:Bar_all_cases}. The plot suggests a positive bias towards class 1, and a negative bias in the labels frequencies in other classes. As confirmed later, the first is due to the large number of FPs for class one, the latter is due to low TP rates for classes 2 and 3. Note that the differences between the frequencies based on VOGN's modal prediction and predictive distribution are irrelevant, while for MCD these are minor and favor predictions based on the predictive density. In the following, we will focus only on results resulting from predictive probabilities. From the analyses in \ref{app:other_analyses} we find that MCD alignment with the sample frequencies is not indicative of a genuine satisfactory performance: for class 1 (classes 2 or 3)  this arises from a lower (comparable) TPR and comparable (lower) FPR with respect to the other optimizers. See e.g. \figurename ~\ref{fig:perf_by_class} therein.

From Table \ref{tab:perf_by_class} it appears that VOGN and ADAM have quite heterogeneous performances based on the measure and class under consideration. In particular, a conclusion on whether it is advisable to use VOGN or ADAM, in general, cannot be made. Overall we observe a tendency for ADAM to perform better in terms of precision and recall, thus on TPs therein involved. Yet when the two are considered jointly (harmonic mean) the f1-score favors VOGN. VOGN furthermore improves the detection of TNs involved in computing accuracy and of course enables the uncertainty analyses based on the predictive distribution.

\begin{table}[htbp]
        \caption{Performance measures for the single-class classification task. }

    \centering
   \scalebox{0.85}{
\begin{tabular}{lcccccc}
\toprule
      & \multicolumn{3}{c}{Precision} & \multicolumn{3}{c}{Accuracy} \\
      & Class 1 & Class 2 & Class 3 & Class 1 & Class 2 & Class 3 \\
\cmidrule(lr){2-4}\cmidrule(lr){5-7} VOGN  & \textbf{0.789} & 0.721 & 0.699 & 0.795 & 0.876 & 0.876 \\
VOGN (pred.) & 0.789 & 0.722 & 0.700 & \textbf{0.795} & \textbf{0.876} & 0.876 \\
ADAM  & 0.774 & \textbf{0.740} & \textbf{0.789} & 0.789 & 0.868 & \textbf{0.888} \\
MCD   & 0.741 & 0.314 & 0.295 & 0.633 & 0.763 & 0.765 \\
MCD (pred.) & 0.757 & 0.381 & 0.361 & 0.684 & 0.794 & 0.798 \\
SGD   & 0.759 & 0.476 & 0.433 & 0.725 & 0.826 & 0.824 \\
\midrule
      &       &       &       &       &       &  \\
      & \multicolumn{3}{c}{Recall} & \multicolumn{3}{c}{f1-score} \\
      & Class 1 & Class 2 & Class 3 & Class 1 & Class 2 & Class 3 \\
\cmidrule(lr){2-4}\cmidrule(lr){5-7} VOGN  & 0.950 & \textbf{0.436} & 0.391 & 0.862 & 0.544 & 0.501 \\
VOGN (pred.) & 0.950 & 0.436 & 0.391 & \textbf{0.862} & \textbf{0.544} & 0.502 \\
ADAM  & \textbf{0.969} & 0.334 & \textbf{0.407} & 0.861 & 0.460 & \textbf{0.537} \\
MCD   & 0.698 & 0.340 & 0.342 & 0.719 & 0.327 & 0.316 \\
MCD (pred.) & 0.780 & 0.347 & 0.349 & 0.768 & 0.363 & 0.355 \\
SGD   & 0.864 & 0.314 & 0.337 & 0.808 & 0.378 & 0.379 \\
\bottomrule

\end{tabular}
}
    \label{tab:perf_by_class}
\end{table}

\subsection{ROC and Calibration Curves} \label{subsec:ROCandCC} % REVIEWED

For our multi-class classification problem, we consider Receiver Operating Characteristic (ROC) curves for the predicted classes. 
In cases where there are no disparities in the cost of false negatives as opposed to false positives, the ROC is a synthetic measure of the quality of models' prediction, irrespective of the chosen classification threshold. To construct ROC curves we discard ambiguous examples by thresholding each validation input's soft-max output and mark the remaining test examples as correctly or incorrectly classified, from which TRP and FPR rates are computed. We apply following thresholds $\{0.05,0.1,0.15,\dots,1\}$. 

\begin{figure}[htbp]
    \centering
    %\scalebox{0.90}{
    \includegraphics[trim={1.3cm 0.1cm 1cm 0.5cm},clip,width=\linewidth]{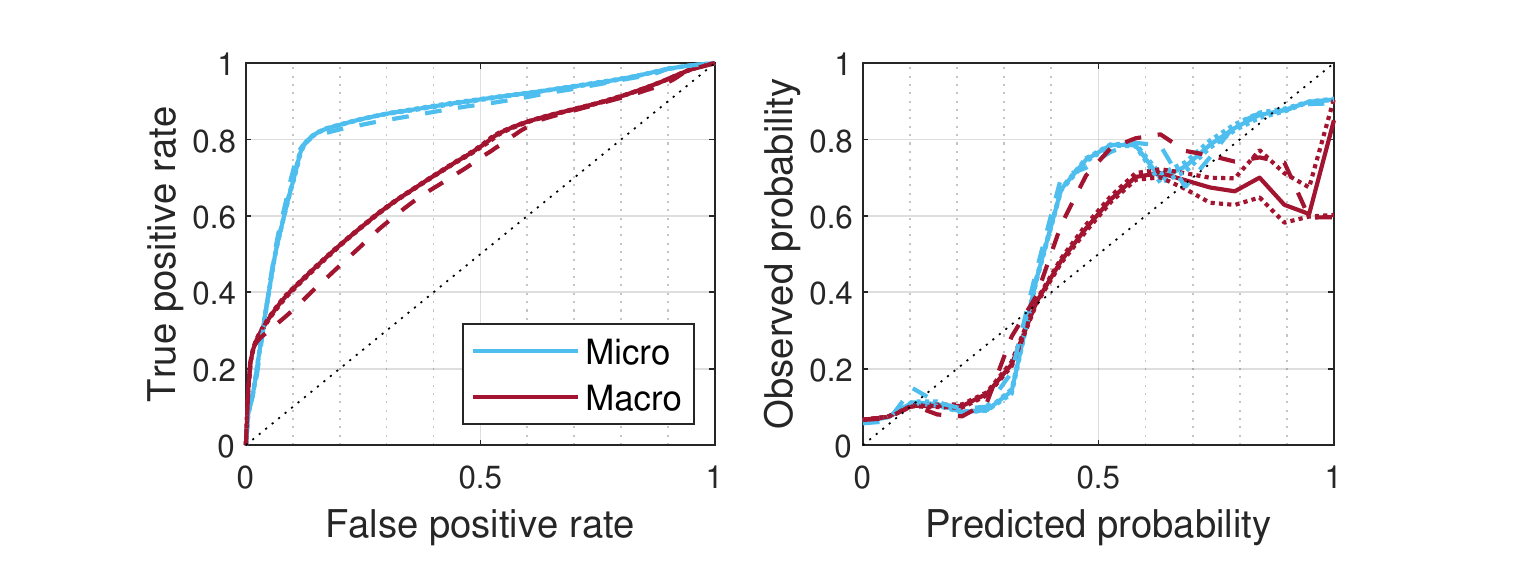}
    %}
    \caption{ROC Curves (left panel) and Calibration curves (right panel). Solid lines: VOGN (predictive), dashed lines: ADAM, dotted lines: 95\% region for VOGN's Calibration curves (sample-by-sample). Legends in the ROC panel apply to CCs as well.}
    \label{fig:ROC_CC_mutliclass}
\end{figure}

\figurename ~\ref{fig:ROC_CC_mutliclass} depicts ROC curves computed from micro and macro FPR and TRP rates for both VOGN and ADAM. For VOGN the figure includes the 95\% interval extracted from the TRP variation across the forecast samples along with the main solid line based on the predictive distribution. 
The multi-class micro and macro averages for VOGN's curves are dominating. 
This indicates that larger predicted scores are increasingly more tightly associated with TP than FP, for VOGN more than for ADAM, and that across the whole FPR domain scores implied by VOGN are more conclusive (in terms of TPs) for the true label.

A commonly reported measure is the FPR at 95\% TPR, which can be interpreted as the probability that a negative example is misclassified as positive when the true positive rate (TPR) is as high as 95\%: for macro-averages we compute 88\% and 90\%, and for micro-averages 76\% and 77\%, for VOGN's forecasts based on the predictive distribution and ADAM respectively.
To assess how well-calibrated a model is, CC compare how well the true class frequency determined by a classifier are calibrated to the true frequency of the positive class, for binned predictions (we take 20 bins). The CC curve of a perfectly calibrated model would lie on the diagonal curve, while over-confident predictions would generally result in CC above the diagonal.

CCs in the left panel of \figurename ~\ref{fig:ROC_CC_mutliclass} underline a comparable performance on micro- and macro- averages. A remarkable S-shape occurs at lower predicted probabilities. This means that the overall satisfactory statistics in Table \ref{tab:AUROC} arise from a balance between the non-ideal scenario where the models are either too much overconfident (predicted probabilities around 20\%) and too much underconfident (around 50\%). 
That is, the underlying scores shift from associating too little probability to the true label to way too much. At high scores, both VOGN's and ADAM's micro-average is quite aligned to the diagonal, yet macro-averages are over-confident suggesting high FPs for the dominant class 1.

\begin{figure}[htbp]
    \centering
    \includegraphics[trim={1.3cm 0.1cm 1cm 0.5cm},clip,width=\linewidth]{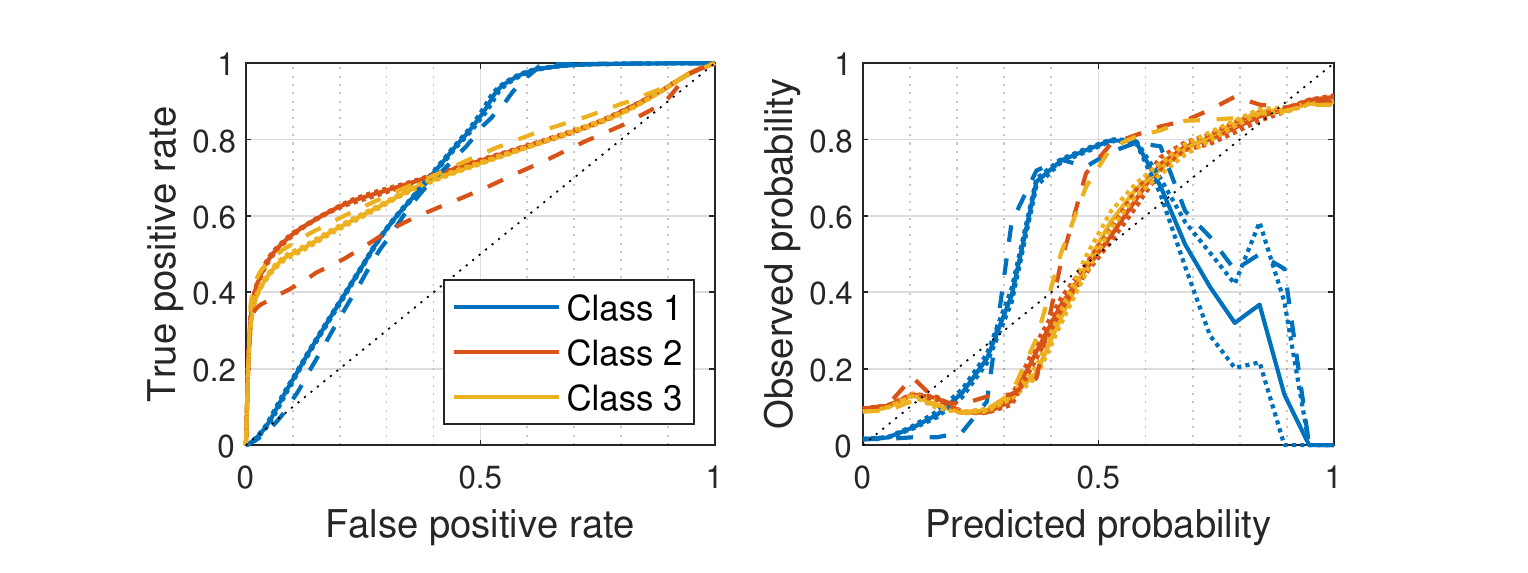}
    %}
    \caption{ROC Curves (left panel) and Calibration curves (right panel). Solid lines: VOGN, dashed lines: ADAM, dotted lines: 95\% region for VOGN's Calibration curves. Legends in the ROC panel apply to CCs as well.}
    \label{fig:ROC_CC_single_class}
\end{figure}

ROC and CCs plots for the single-class task are found in \figurename ~\ref{fig:ROC_CC_single_class}. From the ROC panel, we observe that VOGN outperforms ADAM in classifying labels of classes 1 and 2, and it has a slightly lower performance on class 3. As for Table \ref{tab:perf_by_class} ADAM's metrics are higher than VOGN's for class 3, determining improved TPRs. CCs for classes 1 and 2 are quite satisfactory, and the same comment applies as for the CCs in \figurename ~\ref{fig:ROC_CC_mutliclass}. Remarkable is however the U-shape of the curves for class 1: high class-1 probabilities are overconfident and misleading as there are no samples in class 1 at all when models' probabilities for class 1 are about 1 (confirming the inference from micro- and macro- CCs in  \figurename ~\ref{fig:ROC_CC_mutliclass}).
Aligned with the discussion in Section \ref{subsubsec:model_learning}, models are truly learning the classification of classes 2 and 3. For samples in classes 2 and 3 which however do not display typical class 2 or 3 features, scores associated with classes 2 and 3 are about zero, and all the probability mass is allocated on class 1. In fact, out of the (only) 20 class-1 probabilities higher than 0.75, the 75\% of them correspond to FNs for classes 2 or 3.
This might be indicative of inadequacy in networks' architecture in uncovering deeper patterns in the data that could address class 2 and 3 classification, or non-stationarity elements of true and atypical surprise not observed in the training set or perhaps not learnable at all due to their randomness.

Table \ref{tab:AUROC} reports the Area Under the ROC (AUROC), the Expected Calibration error (ECE), and the L2-norm Distance (ECD) between the CCs and the diagonal line and the CCs. High AUROC, small ECD, and small (in absolute value) ECE are preferred.
Results are aligned with the earlier plots and confirm the above comments. The low ECEs for MCD are not to be interpreted as evidence of improved calibration, as they arise from rather symmetric S-shaped CCs, that however largely deviate from the diagonal (see ECDs). 

\begin{table}
    \caption{Measures related to ROC curves and CCs.}

    \centering
    \scalebox{0.85}{
    \begin{tabular}{lccccc}
\toprule
      & \multicolumn{3}{c}{Single-class task} & \multicolumn{2}{c}{Multi-class task} \\
      & Class 1 & Class 2 & Class 3 & Micro & Macro \\
\cmidrule(lr){2-4}\cmidrule(lr){5-6}\textit{\textbf{Area under the ROC curve}} &       &       &       &       &  \\
VOGN (pred.) & \textbf{0.716} & \textbf{0.739} & 0.722 & \textbf{0.858} & \textbf{0.726} \\
ADAM  & 0.697 & 0.665 & \textbf{0.742} & 0.851 & 0.702 \\
MCD (pred.) & 0.672 & 0.649 & 0.657 & 0.770 & 0.659 \\
SGD   & 0.691 & 0.660 & 0.656 & 0.790 & 0.669 \\
\textit{\textbf{Expected calibration error}} &       &       &       &       &  \\
VOGN (pred.) & -0.107 & -0.014 & \textbf{-0.016} & 0.035 & -0.046 \\
ADAM  & -0.104 & 0.043 & 0.033 & 0.040 & \textbf{-0.009} \\
MCD (pred.) & \textbf{-0.051} & \textbf{-0.016} & -0.044 & \textbf{0.021} & -0.039 \\
SGD   & 0.153 & -0.081 & -0.072 & -0.021 & -0.032 \\
\textit{\textbf{Expected calibration distance}} &       &       &       &       &  \\
VOGN (pred.) & 0.144 & \textbf{0.008} & \textbf{0.009} & \textbf{0.018} & \textbf{0.018} \\
ADAM  & 0.146 & 0.021 & 0.018 & 0.018 & 0.030 \\
MCD (pred.) & 0.181 & 0.028 & 0.012 & 0.019 & 0.023 \\
SGD   & \textbf{0.039} & 0.028 & 0.027 & 0.024 & 0.018 \\
\bottomrule
\end{tabular}
}
\label{tab:AUROC}
\end{table}

\section{Conclusion} % REVIEWED
We proposed a first econometric time-series application with Bayesian Neural Networks. Our task focuses on predicting the direction of mid-price changes in modern limit-order book markets. By utilizing a state-of-the-art optimizer for Bayesian learning and adopting the suitable TABL capable of fully exploiting the ultra-high frequency and complex multidimensional nature of the data, we obtain promising results showing that Bayesian methods in deep learning are feasible, attractive, and valuable for economic applications.

With a number of detailed analyses, we compare several optimizers on the same forecasting exercise and unveil that the Bayesian VOGN optimizer provides, on a general level, the best performance metrics on both multi-class and single-class classification tasks. Yet, VOGN's performance is comparable with the well-known and reliable optimization scheme provided by ADAM. At the same time, Monte Carlo dropout and Stochastic gradient descent methods do not seem to be suitable for the task under analysis. 
Furthermore, we extensively interpret and discuss the results, grasping important insights into the model's learning and decision process. 
The unique feature of Bayesian methods is that of providing posterior and predictive distributions, leading to estimates of the uncertainties associated with the forecasts. The paper discusses how to use and interpret predictive probabilities, providing insights into their implication in the decision process. Following our analysis, and besides promoting further research and applications involving Bayesian deep learning methods, future research might explore to which extent posterior probabilities lead to better uncertainty-informed trades, e.g., by applying and comparing Bayesian and non-Bayesian models for constructing actionable trading strategies, verified with robust back-testing procedures.

\section{Acknowledgments}
This project has received funding from the European Union’s Horizon 2020 research and innovation programme under the Marie Sk\l odowska-Curie grant agreement No. 890690, and the Independent Research Fund Denmark project DISPA (project No. 9041-00004).

\bibliographystyle{IEEEtran}
\bibliography{Bibliography}

\begin{thebibliography}{10}
\providecommand{\url}[1]{#1}
\csname url@samestyle\endcsname
\providecommand{\newblock}{\relax}
\providecommand{\bibinfo}[2]{#2}
\providecommand{\BIBentrySTDinterwordspacing}{\spaceskip=0pt\relax}
\providecommand{\BIBentryALTinterwordstretchfactor}{4}
\providecommand{\BIBentryALTinterwordspacing}{\spaceskip=\fontdimen2\font plus
\BIBentryALTinterwordstretchfactor\fontdimen3\font minus
  \fontdimen4\font\relax}
\providecommand{\BIBforeignlanguage}[2]{{%
\expandafter\ifx\csname l@#1\endcsname\relax
\typeout{** WARNING: IEEEtran.bst: No hyphenation pattern has been}%
\typeout{** loaded for the language `#1'. Using the pattern for}%
\typeout{** the default language instead.}%
\else
\language=\csname l@#1\endcsname
\fi
#2}}
\providecommand{\BIBdecl}{\relax}
\BIBdecl

\bibitem{murphy_machine_2012}
K.~P. Murphy, \emph{Machine learning - a probabilistic perspective}, ser.
  Adaptive computation and machine learning series.\hskip 1em plus 0.5em minus
  0.4em\relax {MIT} Press, 2012.

\bibitem{gal_dropout_2016}
Y.~Gal and Z.~Ghahramani, ``Dropout as a bayesian approximation: Representing
  model uncertainty in deep learning,'' in \emph{33\textsuperscript{rd}
  International Conference on Machine Learning}, vol.~48, 2016, pp. 1050--1059.

\bibitem{frisch1933}
F.~Ragnar, ``Editor's note,'' \emph{Econometrica}, vol.~1, no.~1, pp. 1--4,
  1933.

\bibitem{mullainathan2017machine}
S.~Mullainathan and J.~Spiess, ``Machine learning: an applied econometric
  approach,'' \emph{Journal of Economic Perspectives}, vol.~31, no.~2, pp.
  87--106, 2017.

\bibitem{dixon2020machine}
M.~F. Dixon, I.~Halperin, and P.~Bilokon, \emph{Machine learning in
  Finance}.\hskip 1em plus 0.5em minus 0.4em\relax Springer, 2020, vol. 1170.

\bibitem{varian2014big}
H.~R. Varian, ``Big data: New tricks for econometrics,'' \emph{Journal of
  Economic Perspectives}, vol.~28, no.~2, pp. 3--28, 2014.

\bibitem{salinas2020deepar}
D.~Salinas, V.~Flunkert, J.~Gasthaus, and T.~Januschowski, ``Deepar:
  Probabilistic forecasting with autoregressive recurrent networks,''
  \emph{International Journal of Forecasting}, vol.~36, no.~3, pp. 1181--1191,
  2020.

\bibitem{makridakis2009forecasting}
S.~Makridakis, R.~M. Hogarth, and A.~Gaba, ``Forecasting and uncertainty in the
  economic and business world,'' \emph{International Journal of Forecasting},
  vol.~25, no.~4, pp. 794--812, 2009.

\bibitem{kingma_auto_2014}
D.~P. Kingma and M.~Welling, ``Auto-encoding variational bayes,'' in
  \emph{2\textsuperscript{nd} International Conference on Learning
  Representations, {ICLR}}, Y.~Bengio and Y.~LeCun, Eds., 2014, pp. 1--14.

\bibitem{osawa_practical_2019}
K.~Osawa, S.~Swaroop, M.~E. Khan, A.~Jain, R.~Eschenhagen, R.~E. Turner, and
  R.~Yokota, ``Practical deep learning with bayesian principles,'' in
  \emph{Advances in Neural Information Processing Systems}, vol.~32, 2019, pp.
  1--13.

\bibitem{blundell_weight_2015}
C.~Blundell, J.~Cornebise, K.~Kavukcuoglu, and D.~Wierstra, ``Weight
  uncertainty in neural network,'' in \emph{32\textsuperscript{nd}
  International Conference on Machine Learning}.\hskip 1em plus 0.5em minus
  0.4em\relax PMLR, 2015, pp. 1613--1622.

\bibitem{geweke2010comparing}
J.~Geweke and G.~Amisano, ``Comparing and evaluating bayesian predictive
  distributions of asset returns,'' \emph{International Journal of
  Forecasting}, vol.~26, no.~2, pp. 216--230, 2010.

\bibitem{krizhevsky_imagenet_2012}
A.~Krizhevsky, I.~Sutskever, and G.~E. Hinton, ``Imagenet classification with
  deep convolutional neural networks,'' \emph{Advances in neural information
  processing systems}, vol.~25, 2012.

\bibitem{simonyan_very_2015}
K.~Simonyan and A.~Zisserman, ``Very deep convolutional networks for
  large-scale image recognition,'' in \emph{3\textsuperscript{rd} International
  Conference on Learning Representations}, 2015, pp. 1--14.

\bibitem{girshick_rich_2014}
R.~Girshick, J.~Donahue, T.~Darrell, and J.~Malik, ``Rich feature hierarchies
  for accurate object detection and semantic segmentation,'' in \emph{IEEE
  conference on computer vision and pattern recognition}, 2014, pp. 580--587.

\bibitem{ren_faster_2015}
S.~Ren, K.~He, R.~Girshick, and J.~Sun, ``Faster r-cnn: Towards real-time
  object detection with region proposal networks,'' \emph{Advances in neural
  information processing systems}, vol.~28, 2015.

\bibitem{collobert_unified_2008}
R.~Collobert and J.~Weston, ``A unified architecture for natural language
  processing: deep neural networks with multitask learning,'' in
  \emph{25\textsuperscript{th} International Conference on Machine Learning},
  2008, pp. 160--167.

\bibitem{goldberg_neural_2017}
Y.~Goldberg, \emph{Neural Network Methods for Natural Language
  Processing}.\hskip 1em plus 0.5em minus 0.4em\relax Morgan \& Claypool
  Publishers, 2017.

\bibitem{mohamed_acoustic_2012}
A.-r. Mohamed, G.~E. Dahl, and G.~E. Hinton, ``Acoustic modeling using deep
  belief networks,'' \emph{IEEE Transactions on Speech and Audio Processing},
  vol.~20, no.~1, pp. 14--22, 2012.

\bibitem{dahl_context-dependent_2012}
G.~E. Dahl, D.~Yu, L.~Deng, and A.~Acero, ``Context-dependent pre-trained deep
  neural networks for large-vocabulary speech recognition,'' \emph{IEEE
  Transactions on Speech and Audio Processing}, vol.~20, no.~1, pp. 30--42,
  2012.

\bibitem{cybenko_approximation_1989}
G.~Cybenko, ``Approximation by superpositions of a sigmoidal function,''
  \emph{Math. Control. Signals Syst}, vol.~2, no.~4, pp. 303--314, 1989.

\bibitem{Lu2017expressive}
Z.~Lu, H.~Pu, F.~Wang, Z.~Hu, and L.~Wang, ``The expressive power of neural
  networks: A view from the width,'' in \emph{Advances in Neural Information
  Processing Systems}, vol.~30, 2017, pp. 1--9.

\bibitem{Hanin2019universal}
B.~Hanin, ``Universal function approximation by deep neural nets with bounded
  width and relu activations,'' \emph{Mathematics}, vol.~7, no.~10, p. 992,
  2019.

\bibitem{mcnelis_neural_2005}
P.~D. McNelis, \emph{Neural networks in finance: gaining predictive edge in the
  market}, ser. Academic Press advanced finance series.\hskip 1em plus 0.5em
  minus 0.4em\relax Elsevier Academic Press, 2005.

\bibitem{kuan_artificial_1994}
C.-M. Kuan and H.~White, ``Artificial neural networks: an econometric
  perspective,'' \emph{Econometric Reviews}, vol.~13, no.~1, pp. 1--91, 1994.

\bibitem{terasvirta_linear_2005}
T.~Teräsvirta, D.~van Dijk, and M.~C. Medeiros, ``Linear models, smooth
  transition autoregressions, and neural networks for forecasting macroeconomic
  time series: A re-examination,'' \emph{Nonlinearities, Business Cycles and
  Forecasting}, vol.~21, no.~4, pp. 755--774, 2005.

\bibitem{m_qi_trend_2008}
M.~Qi and G.~P. Zhang, ``Trend time–series modeling and forecasting with
  neural networks,'' \emph{IEEE Transactions on Neural Networks}, vol.~19,
  no.~5, pp. 808--816, 2008.

\bibitem{hewamalage2021recurrent}
H.~Hewamalage, C.~Bergmeir, and K.~Bandara, ``Recurrent neural networks for
  time series forecasting: Current status and future directions,''
  \emph{International Journal of Forecasting}, vol.~37, no.~1, pp. 388--427,
  2021.

\bibitem{cenesizoglu_effects_2014}
T.~Cenesizoglu, G.~Dionne, and X.~Zhou, ``Effects of the limit order book on
  price dynamics,'' \emph{Microeconomics: General Equilibrium \& Disequilibrium
  Models of Financial Markets eJournal}, 2014.

\bibitem{kercheval_modelling_2015}
A.~N. Kercheval and Y.~Zhang, ``Modelling high-frequency limit order book
  dynamics with support vector machines,'' \emph{Quantitative Finance},
  vol.~15, no.~8, pp. 1315--1329, 2015.

\bibitem{ntakaris_feature_2019}
A.~Ntakaris, G.~Mirone, J.~Kanniainen, M.~Gabbouj, and A.~Iosifidis, ``Feature
  engineering for mid-price prediction with deep learning,'' \emph{IEEE
  Access}, vol.~7, pp. 82\,390--82\,412, 2019.

\bibitem{ntakaris_benchmark_2018}
A.~Ntakaris, M.~Magris, J.~Kanniainen, M.~Gabbouj, and A.~Iosifidis,
  ``Benchmark dataset for mid-price forecasting of limit order book data with
  machine learning methods,'' \emph{Journal of Forecasting}, vol.~37, no.~8,
  pp. 852--866, 2018.

\bibitem{dixon_sequence_2018}
M.~Dixon, ``Sequence classification of the limit order book using recurrent
  neural networks,'' \emph{Journal of Computational Science}, vol.~24, pp.
  277--286, 2018.

\bibitem{tsantekidis_forecasting_2017}
A.~Tsantekidis, N.~Passalis, A.~Tefas, J.~Kanniainen, M.~Gabbouj, and
  A.~Iosifidis, ``Forecasting stock prices from the limit order book using
  convolutional neural networks,'' in \emph{19\textsuperscript{th} {IEEE}
  Conference on Business Informatics}, 2017, pp. 7--12.

\bibitem{zhang2019deeplob}
Z.~Zhang, S.~Zohren, and S.~Roberts, ``Deeplob: Deep convolutional neural
  networks for limit order books,'' \emph{IEEE Transactions on Signal
  Processing}, vol.~67, no.~11, pp. 3001--3012, 2019.

\bibitem{passalis2019adaptive}
N.~Passalis, A.~Tefas, J.~Kanniainen, M.~Gabbouj, and A.~Iosifidis, ``Deep
  adaptive input normalization for time series forecasting,'' \emph{IEEE
  Transactions on Neural Networks and Learning Systems}, vol.~31, no.~9, pp.
  3760--3765, 2019.

\bibitem{tsantekidis_using_2020}
A.~Tsantekidis, N.~Passalis, A.~Tefas, J.~Kanniainen, M.~Gabbouj, and
  A.~Iosifidis, ``Using deep learning for price prediction by exploiting
  stationary limit order book features,'' \emph{Applied Soft Computing},
  vol.~93, p. 106401, 2020.

\bibitem{passalis2017time}
N.~Passalis, A.~Tsantekidis, A.~Tefas, J.~Kanniainen, M.~Gabbouj, and
  A.~Iosifidis, ``Time-series classification using neural bag-of-features,'' in
  \emph{European Signal Processing Conference}, 2017, pp. 301--305.

\bibitem{passalis2018temporal}
N.~Passalis, A.~Tefas, J.~Kanniainen, M.~Gabbouj, and A.~Iosifidis, ``Temporal
  bag-of-features learning for predicting mid price movements using high
  frequency limit order book data,'' \emph{IEEE Transactions on Emerging Topics
  in Computational Intelligence}, vol.~4, no.~6, pp. 774--785, 2018.

\bibitem{sirignano_deep_2019}
J.~A. Sirignano, ``Deep learning for limit order books,'' \emph{Quantitative
  Finance}, vol.~19, no.~4, pp. 549--570, 2019.

\bibitem{tran2017tensor}
D.~T. Tran, M.~Magris, J.~Kanniainen, M.~Gabbouj, and A.~Iosifidis, ``Tensor
  representation in high-frequency financial data for price change
  prediction,'' in \emph{IEEE Symposium Series on Computational Intelligence},
  2017, pp. 1--7.

\bibitem{tran_temporal_2019}
D.~T. Tran, A.~Iosifidis, J.~Kanniainen, and M.~Gabbouj, ``Temporal
  {Attention}-{Augmented} {Bilinear} {Network} for {Financial} {Time}-{Series}
  {Data} {Analysis},'' \emph{IEEE Transactions on Neural Networks and Learning
  Systems}, vol.~30, no.~5, pp. 1407--1418, 2019.

\bibitem{shabani_multi-head_2022}
M.~Shabani, D.~T. Tran, M.~Magris, J.~Kanniainen, and A.~Iosifidis,
  ``Multi-head temporal attention-augmented bilinear network for financial time
  series prediction,'' in \emph{2022 30th European Signal Processing Conference
  (EUSIPCO)}, 2022, pp. 1487--1491.

\bibitem{goan_bayesian_2020}
E.~Goan and C.~Fookes, \emph{Bayesian Neural Networks: An Introduction and
  Survey}.\hskip 1em plus 0.5em minus 0.4em\relax Springer International
  Publishing, 2020, ch.~3, pp. 45--87.

\bibitem{caruana_intelligible_2015}
R.~Caruana, Y.~Lou, J.~Gehrke, P.~Koch, M.~Sturm, and N.~Elhadad,
  ``Intelligible models for healthcare: Predicting pneumonia risk and hospital
  30-day readmission,'' in \emph{of the 21\textsuperscript{st} International
  Conference on Knowledge Discovery and Data Mining}, 2015, pp. 1721--1730.

\bibitem{holzinger_what_2017}
A.~Holzinger, C.~Biemann, C.~S. Pattichis, and D.~B. Kell, ``What do we need to
  build explainable ai systems for the medical domain?''
  \emph{arXiv:1712.09923}, 2017.

\bibitem{vu_shared_2018}
M.-A.~T. Vu, T.~Adalı, D.~Ba, G.~Buzsáki, D.~Carlson, K.~Heller, C.~Liston,
  C.~Rudin, V.~S. Sohal, A.~S. Widge, H.~S. Mayberg, G.~Sapiro, and K.~Dzirasa,
  ``A shared vision for machine learning in neuroscience,'' \emph{Journal of
  Neuroscience}, vol.~38, no.~7, pp. 1601--1607, 2018.

\bibitem{holzinger_causability_2019}
A.~Holzinger, G.~Langs, H.~Denk, K.~Zatloukal, and H.~Müller, ``Causability
  and explainability of artificial intelligence in medicine,'' \emph{WIREs Data
  Mining and Knowledge Discovery}, vol.~9, no.~4, p. e1312, 2019.

\bibitem{mackay_probable_1995}
D.~J.~C. Mackay, ``Probable networks and plausible predictions - a review of
  practical bayesian methods for supervised neural networks,'' \emph{Network:
  Computation in Neural Systems}, vol.~6, no.~3, pp. 469--505, 1995.

\bibitem{lampinen_bayesian_2001}
J.~Lampinen and A.~Vehtari, ``Bayesian approach for neural networks—review
  and case studies,'' \emph{Neural Networks}, vol.~14, no.~3, pp. 257--274,
  2001.

\bibitem{jospin_hands_2020}
L.~V. Jospin, W.~Buntine, F.~Boussaid, H.~Laga, and M.~Bennamoun, ``Hands-on
  bayesian neural networks--a tutorial for deep learning users,''
  \emph{arXiv:2007.06823}, 2020.

\bibitem{magris2023}
M.~Magris and A.~Iosifidis, ``Bayesian learning for neural networks: an
  algorithmic survey,'' \emph{arXiv:2211.11865}, 2022.

\bibitem{mbuvha_automatic_2019}
R.~Mbuvha, I.~Boulkaibet, and T.~Marwala, ``Automatic relevance determination
  bayesian neural networks for credit card default modelling,''
  \emph{arXiv:1906.06382}, 2019.

\bibitem{chandra_bayesian_2021}
R.~Chandra and Y.~He, ``Bayesian neural networks for stock price forecasting
  before and during covid-19 pandemic,'' \emph{PLOS One}, vol.~16, no.~7, pp.
  1--32, 07 2021.

\bibitem{vahidinasab_bayesian_2008}
V.~Vahidinasab and S.~Jadid, ``Bayesian neural network model to predict
  day-ahead electricity prices,'' \emph{European Transactions on Electrical
  Power}, vol.~20, pp. 231--246, 2008.

\bibitem{ghayekhloo_combination_2019}
M.~Ghayekhloo, R.~Azimi, M.~Ghofrani, M.~B. Menhaj, and E.~Shekari, ``A
  combination approach based on a novel data clustering method and bayesian
  recurrent neural network for day-ahead price forecasting of electricity
  markets,'' \emph{Electric Power Systems Research}, 2019.

\bibitem{jang_empirical_2018}
H.~Jang and J.~Lee, ``An empirical study on modeling and prediction of bitcoin
  prices with bayesian neural networks based on blockchain information,''
  \emph{{IEEE} Access}, vol.~6, pp. 5427--5437, 2018.

\bibitem{skabar_direction-change_2009}
A.~A. Skabar, \emph{Direction-of-Change Financial Time Series Forecasting Using
  Neural Networks: A Bayesian Approach}.\hskip 1em plus 0.5em minus 0.4em\relax
  Springer, 2009.

\bibitem{blei2017variational}
D.~M. Blei, A.~Kucukelbir, and J.~D. McAuliffe, ``Variational inference: A
  review for statisticians,'' \emph{Journal of the American statistical
  Association}, vol. 112, no. 518, pp. 859--877, 2017.

\bibitem{tran_practical_2021}
M.-N. Tran, T.-N. Nguyen, and V.-H. Dao, ``A practical tutorial on variational
  bayes,'' \emph{arXiv:2103.01327}, 2021.

\bibitem{gunawan2021variational}
D.~Gunawan, R.~Kohn, and D.~Nott, ``Variational bayes approximation of factor
  stochastic volatility models,'' \emph{International Journal of Forecasting},
  vol.~37, no.~4, pp. 1355--1375, 2021.

\bibitem{gefang2022}
D.~Gefang, G.~Koop, and A.~Poon, ``Forecasting using variational bayesian
  inference in large vector autoregressions with hierarchical shrinkage,''
  \emph{International Journal of Forecasting}, 2022.

\bibitem{khan_2018_fast}
M.~E. Khan and D.~Nielsen, ``Fast yet simple natural-gradient descent for
  variational inference in complex models,'' in \emph{International Symposium
  on Information Theory and Its Applications}, 2018, pp. 31--35.

\bibitem{khan_2017_conjugate}
M.~E. Khan and W.~Lin, ``Conjugate-computation variational inference:
  converting variational inference in non-conjugate models to inferences in
  conjugate models,'' in \emph{20\textsuperscript{th} International Conference
  on Artificial Intelligence and Statistics}, 2017, pp. 878--887.

\bibitem{blundell2015weight}
C.~Blundell, J.~Cornebise, K.~Kavikcuoglu, and D.~Wierstra, ``Weight
  uncertainty in neural networks,'' in \emph{International Conference on
  Machine Learning}, 2015, pp. 1613--1622.

\bibitem{osawa_2018_torch}
K.~Osawa, ``Pytorch-sso: Scalable second-order methods in pytorch,'' 2019,
  gitHub repository.

\bibitem{cont_stochastic_2010}
R.~Cont, S.~Stoikov, and R.~Talreja, ``A stochastic model for the order book
  dynamics,'' \emph{Operations Research}, vol.~58, no.~3, pp. 549--563, 2010.

\bibitem{tran2021levels}
D.~T. Tran, J.~Kanniainen, and A.~Iosifidis, ``{How informative is the Order
  Book Beyond the Best Levels? Machine Learning Perspective},'' in
  \emph{NeurIPS 2021 Workshop on Machine Learning meets Econometrics}, 2021,
  pp. 1--12.

\bibitem{kingma_adam_2015}
D.~P. Kingma and J.~L. Ba, ``Adam: A method for stochastic optimization,'' in
  \emph{3\textsuperscript{rd} International Conference on Learning
  Representations}, 2015, pp. 1--15.

\end{thebibliography}

\appendix

\section{Performance on individual stocks} \label{app:stock_perf}

All the models are trained in an end-to-end manner over stacked features and labels corresponding to five stocks. As a sanity check, we report in Table \ref{tab:perf_on_stocks} the performance for the multi-class task for each of them. We observe metrics aligned in magnitudes with the overall ones in Table \ref{tab:perf_all}, confirming a qualitative consistency in the data across different stocks, the reliability of the results, and the robustness of the methods. Standard deviations in the metrics are lowest for VOGN, proving a firmer consistency in the results and perhaps a better generalization ability to unseen market data. 

\begin{table*}
    \caption{Performance measures for the multi-class classification task on different stocks. }
    \centering
    %\scalebox{0.75}{
\begin{tabular}{lccccccc}
\toprule
      & Any   & \multicolumn{2}{c}{Precision} & \multicolumn{2}{c}{Recall} & \multicolumn{2}{c}{f1-score} \\
      & Micro & Macro & Weighted & Macro & Weighted & Macro & Weighted \\
\cmidrule(lr){2-2}\cmidrule(lr){3-4}\cmidrule(lr){5-6}\cmidrule(lr){7-8} \multicolumn{8}{c<{\vspace{-0.3cm}}}{}  \\
\multicolumn{8}{c}{\textit{{Stock: Kesko Oyj, ISIN: FI0009000202}}} \\
%VOGN  & \textbf{0.777} & 0.738 & 0.766 & \textbf{0.595} & \textbf{0.777} & \textbf{0.639} & \textbf{0.755} \\
VOGN (pred.) & 0.776 & 0.732 & 0.764 & 0.594 & 0.776 & 0.636 & 0.753 \\
ADAM  & 0.776 & \textbf{0.771} & \textbf{0.774} & 0.574 & 0.776 & 0.624 & 0.746 \\
%MCD   & 0.581 & 0.448 & 0.600 & 0.459 & 0.581 & 0.453 & 0.590 \\
MCD (pred.) & 0.638 & 0.497 & 0.632 & 0.491 & 0.638 & 0.494 & 0.635 \\
SGD   & 0.690 & 0.558 & 0.663 & 0.507 & 0.690 & 0.524 & 0.671 \\[1ex]
\multicolumn{8}{c}{\textit{{Stock: Outokumpu Oyj, ISIN: FI0009002422}}} \\
%VOGN  & 0.743 & 0.663 & 0.730 & 0.590 & 0.743 & 0.616 & 0.729 \\
VOGN (pred.) & \textbf{0.743} & \textbf{0.663} & 0.730 & 0.591 & \textbf{0.743} & \textbf{0.616} & \textbf{0.730} \\
ADAM  & 0.667 & 0.656 & \textbf{0.738} & \textbf{0.602} & 0.667 & 0.595 & 0.685 \\
%MCD   & 0.574 & 0.438 & 0.582 & 0.433 & 0.574 & 0.432 & 0.576 \\
MCD (pred) & 0.607 & 0.469 & 0.600 & 0.447 & 0.607 & 0.451 & 0.600 \\
SGD   & 0.659 & 0.527 & 0.652 & 0.518 & 0.659 & 0.522 & 0.655 \\[1ex]
\multicolumn{8}{c}{\textit{{Stock: Rautaruukki Oyj, ISIN: FI0009003552}}} \\
VOGN  & \textbf{0.748} & \textbf{0.669} & 0.735 & 0.599 & \textbf{0.748} & \textbf{0.624} & \textbf{0.735} \\
VOGN (pred.) & 0.747 & 0.669 & 0.735 & 0.599 & 0.747 & 0.624 & 0.735 \\
ADAM  & 0.675 & 0.662 & \textbf{0.744} & \textbf{0.613} & 0.675 & 0.605 & 0.692 \\
%MCD   & 0.573 & 0.439 & 0.581 & 0.435 & 0.573 & 0.433 & 0.576 \\
MCD (pred) & 0.607 & 0.470 & 0.601 & 0.450 & 0.607 & 0.453 & 0.600 \\
SGD   & 0.663 & 0.534 & 0.658 & 0.527 & 0.663 & 0.530 & 0.660 \\[1ex]
\multicolumn{8}{c}{\textit{{Stock: Sampo Oyj, ISIN: FI0009003305}}} \\
%%VOGN  & \textbf{0.743} & \textbf{0.664} & 0.730 & 0.592 & \textbf{0.743} & \textbf{0.618} & \textbf{0.730} \\
VOGN (pred.) & 0.743 & 0.663 & 0.730 & 0.592 & 0.743 & 0.617 & 0.730 \\
ADAM  & 0.669 & 0.658 & \textbf{0.739} & \textbf{0.605} & 0.669 & 0.598 & 0.686 \\
%MCD   & 0.573 & 0.436 & 0.579 & 0.432 & 0.573 & 0.431 & 0.575 \\
MCD (pred) & 0.608 & 0.467 & 0.599 & 0.446 & 0.608 & 0.451 & 0.600 \\
SGD   & 0.659 & 0.526 & 0.653 & 0.519 & 0.659 & 0.522 & 0.655 \\[1ex]
\multicolumn{8}{c}{\textit{{Stock: W\"{a}rtsil\"{a} Oyj, ISIN: FI0009000727}}} \\
%VOGN  & 0.747 & \textbf{0.666} & 0.735 & 0.600 & 0.747 & 0.624 & 0.736 \\
VOGN (pred.) & \textbf{0.747} & 0.666 & 0.735 & 0.600 & \textbf{0.747} & \textbf{0.625} & \textbf{0.736} \\
ADAM  & 0.675 & 0.661 & \textbf{0.743} & \textbf{0.613} & 0.675 & 0.604 & 0.692 \\
%MCD   & 0.579 & 0.442 & 0.587 & 0.439 & 0.579 & 0.438 & 0.582 \\
MCD (pred) & 0.615 & 0.476 & 0.608 & 0.457 & 0.615 & 0.461 & 0.609 \\
SGD   & 0.663 & 0.532 & 0.659 & 0.527 & 0.663 & 0.529 & 0.661 \\[1ex]
\midrule
\multicolumn{8}{c}{\textit{{Standard deviation}}} \\
%VOGN  & 0.014 & 0.033 & 0.015 & 0.004 & 0.014 & 0.009 & 0.010 \\
VOGN (pred.) & \textbf{0.014} & \textbf{0.030} & \textbf{0.014} & \textbf{0.004} & \textbf{0.014} & \textbf{0.008} & \textbf{0.010} \\
ADAM  & 0.047 & 0.050 & 0.015 & 0.016 & 0.047 & 0.011 & 0.026 \\
\bottomrule
\end{tabular}%
%}

    \label{tab:perf_on_stocks}
\end{table*}

\section{Further details on the single-class classification task}\label{app:other_analyses} 
\begin{figure}
    \centering
    %\scalebox{0.99}{
    \includegraphics[trim={0.5cm 1.2cm 0.5cm 0.5cm},clip,width=\columnwidth]{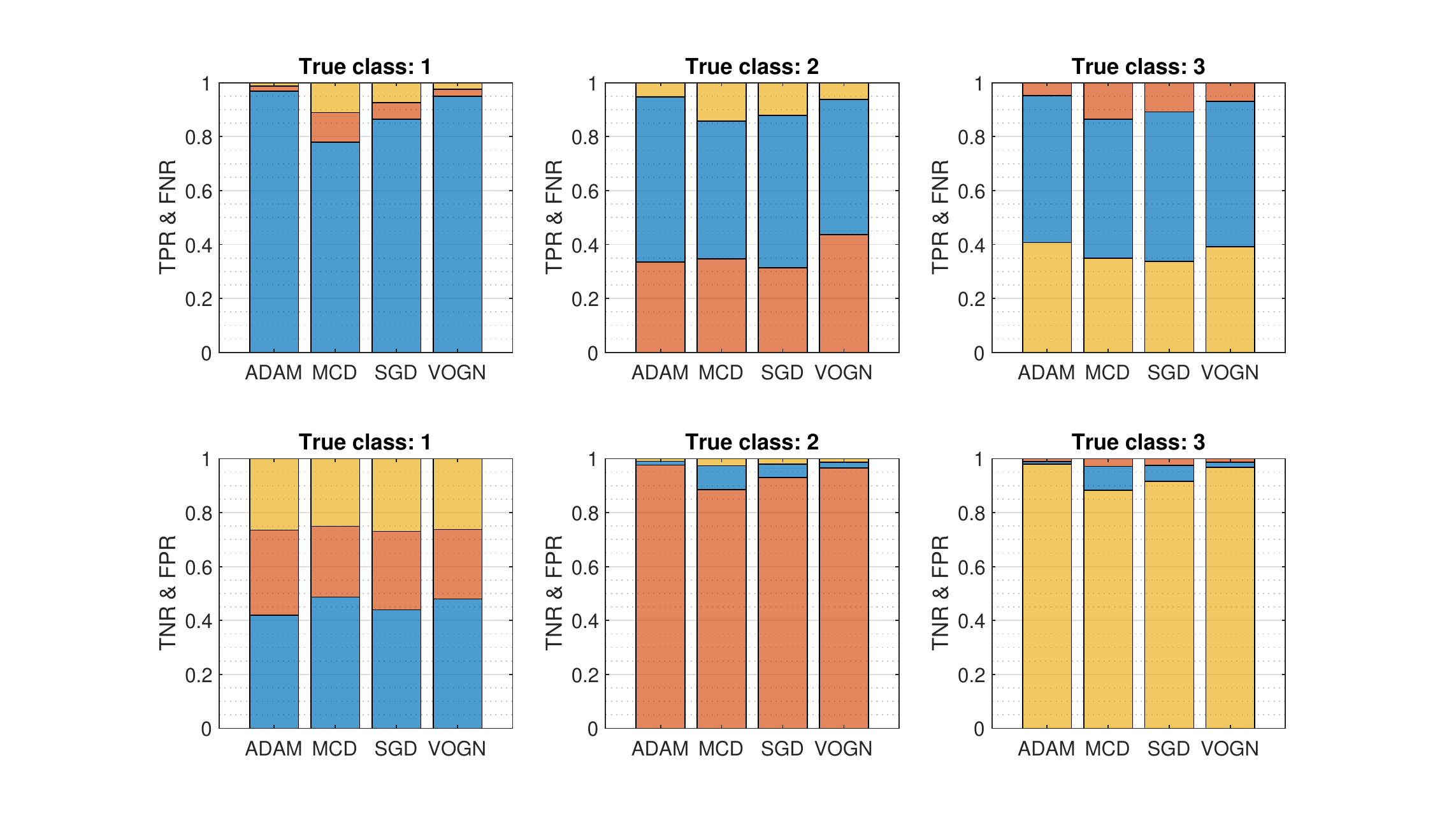}
    %}
    \caption{Top row: True Positive Rates (TPRs) and distribution of the false negatives. TRPs correspond to the heights of the lowest bars, while False Negative Rates (FNRs) are extracted as their complement to one. Bottom row: True Negative Rates (TNR) and  distribution of the false positives. FNRs correspond to the heights of the lowest bars, while False Positive Rates (FPRs) are extracted as their complement to one. Forecasts for VOGN and MCD are based on the predictive distribution. Classes one, two, three are respectively denoted by blue, red and yellow colors.}
    \label{fig:perf_by_class}
\end{figure}

A number of further considerations can be drawn by analyzing the details of correct and miss-correct assignments for the single-class classification task.
The top-left panel in \figurename ~\ref{fig:perf_by_class} displays a slightly higher TPR rate for ADAM than for VOGN. For all the optimizers, FNs are equally distributed across classes 2 and 3, suggesting that miss-classifications of stationary price movements are due to patterns in the features that are truly atypical, neither representative of class 2 nor 3. Whereas TPRs for class 1 are generally overwhelming with respect to FNRs, the opposite holds for classes 2 and 3. For all the true labels in class 2 or 3, only 35\% of them are detected in such classes (TPs), while more than 50\% are classified as class 1 (FPs). The small fraction of FPs for classes 3 and 2 under the true labels being 2 and 3, underlines that the model confuses price increases (decreases) with stationary prices but not with price decreases (increases). On the other hand, in the bottom row of \ ~\ref{fig:perf_by_class}, we find that TNRs for classes 2 and 3 are very high for all the models, indicating that the models unveil patterns in the features that are truly indicative of classes 2 and 3, that when not detected lead to high TNRs. For class 1, however, TNRs are skimpy, and FPs are equally distributed across classes 2 and 3, underlying the difficulty the model has in detecting features and patterns in the data truly indicative of the stationary-price case. Along with the observations in Subsection \ref{subsubsec:model_learning}, this provides further evidence that the models truly learn a classification rule for upward and downward price movements only. 

A relevant metric for actionable trading decisions is the False Discovery Rate (FDR). FDR indicates the fraction of false discoveries (FP) over the positives (FP and TP), approximating the probability that a foretasted price direction is a FP. FDR  quantifies the risk of undertaking a trading decision (e.g., placing a sell order based on a price-decrease forecast) based on a signal that turns out to be false (price increases). We observe that for all the optimizers the FDR is about 30\% in all three classes, and that for VOGN and ADAM the difference is always well-beyond 1\%. This corresponds to a great achievement upon a random classifier (50\% FDR), yet for business operations, it still represents a substantial risk. FDR is an aggregate measure: for a given example, labels' uncertainties are captured by predictive probabilities.

Lastly, we investigate whether VOGN's predictive distribution is capable of quantifying different uncertainty levels for correctly and miss-classified labels. Indeed, a considerable difference in predictive probabilities between TP and FP as much as TN and FN would be desirable. 
Low uncertainties associated with, e.g., TPs or TNs, would certainly indicate that the predictive distribution is, in fact, well-calibrated, being confident on the assignments that eventually turn out to be correct.

\begin{table}[H]
    \centering
        \caption{VOGN's predictive probabilities across correctly and miss-classified samples for the class of maximum probability  and for the true class. I.e. $\hat{p}^{(1)}_i$ and $\hat{p}_{ic}$ with $c$ being the true label class, respectively }
    \scalebox{0.9}{
\begin{tabular}{lcccccc}
\toprule
 & \multicolumn{3}{c}{$\hat{p}^{(1)}_i$} & \multicolumn{3}{c}{$\hat{p}_{ic}$} \\
 True class:     & Class 1 & Class 2 & Class 3 & Class 1 & Class 2 & Class 3 \\
\cmidrule(lr){2-4}\cmidrule(lr){5-7}TP    & 0.495 & \textbf{0.808} & \textbf{0.809} & 0.495 & \textbf{0.808} & \textbf{0.809} \\
FN    & 0.497 & 0.513 & 0.524 & 0.324 & 0.253 & 0.245 \\
FP    & 0.497 & 0.579 & 0.566 & \textbf{0.497} & 0.579 & 0.566 \\
TN    & \textbf{0.793} & 0.520 & 0.523 & 0.159 & 0.236 & 0.226 \\
\bottomrule
\end{tabular}
}
\label{tab:pred_single_class}
\end{table}

The first three columns in Table \ref{tab:pred_single_class} refer to the predictive distribution of the class of maximum probability $\hat{p}^{(1)}_i$, i.e., the class one would take as a forecast in an actual forecasting exercise. As desirable, TPs for classes 2 and 3 correspond to the predictive probabilities, thus to the lowest uncertainties. However, predictive probabilities are comparable for FNs and TNs and slightly higher for FP. That is, low levels of uncertainties can be safely associated with TPs, yet no insight can be grasped on FN, FP, and TN. Enforcing the observations in Section \ref{subsubsec:model_learning}, high scores in class 1 are associated with TN, indicating that the uncertainty in class 1 is low when actual forecasts are in classes 2 or 3. The last three columns in Table \ref{tab:pred_single_class} refer to the predictive distribution over the true class. This information is clearly unavailable in real settings but useful for model back-testing. Across all the classes, high probabilities are always associated with TPs (desirable), lowest probabilities with TN (would be desirable to observe high values), and about 50\% of the predictive probabilities to FP (indicating noteworthy confidence in forecasts that are indeed miss-classified).

\end{document}